\documentclass[aps,prxquantum,reprint,superscriptaddress,nofootinbib]{revtex4-2}
\pdfoutput=1
\usepackage[utf8]{inputenc}
\usepackage[english]{babel}
\usepackage[T1]{fontenc}
\usepackage{amsmath}
\usepackage{amsthm}
\newtheorem{theorem}{Theorem}
\usepackage{hyperref}
\usepackage{tikz}
\usepackage{lipsum}
\usepackage{amsmath,amssymb}
\usepackage{graphicx}

\usepackage{pifont}    
\usepackage{wrapfig}
\usepackage{algorithm}
\usepackage{algorithmic}
\usepackage[toc,page]{appendix}
\usepackage{comment}

\usepackage{adjustbox} 
\usepackage{enumitem}
\usepackage{caption}
\usepackage{booktabs}  
\usepackage{array}     
\usepackage{subcaption}
\usepackage{overpic}
\usepackage[justification=justified,singlelinecheck=false]{caption}

\usepackage{listings}
\usepackage{xcolor}
\usepackage{braket}

\usepackage{amsmath,amssymb}
\usepackage{braket}

\usepackage{tikz}
\usepackage{quantikz} 

\begin{document}


\title{Quantum State Preparation via Schmidt Spectrum Optimisation}

\newcommand{\UWAaff}{Centre for Quantum Information, Simulation and Algorithms,
School of Physics, Mathematics and Computing, The University of Western Australia, Perth, Australia}

\author{Josh Green}
\email{josh.green@uwa.edu.au}
\affiliation{\UWAaff}
\affiliation{Department of Engineering, University of Cambridge, Cambridge, United Kingdom}

\author{Josh Snow}
\email{josh.snow@uwa.edu.au}
\affiliation{\UWAaff}

\author{Jingbo Wang}
\email{jingbo.wang@uwa.edu.au}
\affiliation{\UWAaff}




\begin{abstract}
Quantum state preparation represents a critical bottleneck for a broad class of quantum algorithms. In this work, we introduce the Schmidt Spectrum Optimisation (SSO) algorithm as an efficient and scalable approach for preparing quantum states described by Matrix Product States (MPS). The SSO algorithm employs a preparation-by-disentangling strategy by optimising circuit layers of two-qubit gates to progressively remove entanglement from a target state. Each circuit layer is computed sequentially and efficiently on a classical computer using tensor network optimisation techniques. Once the target state has been successfully disentangled, a quantum state-preparation circuit is formed by reversing the sequence of optimised disentangling layers. Across benchmarks including random MPS and MPS approximations to the ground-states of local Hamiltonians, we find that the SSO algorithm significantly improves upon prior variational and disentangling-based approaches, highlighting its potential as a scalable framework for quantum state preparation.
\end{abstract}


\maketitle

\section{Introduction}

Quantum state preparation refers to the process of transforming a quantum system from its initial all-zero state into a precisely defined configuration, such as approximations of highly structured many-body systems \cite{huggins2025,dong2022}, image or signal data encoding \cite{Green2025,Jobst2024efficientmps}, or discretised functions \cite{Rosenkranz2025quantumstate,Jaderberg2025,iaconis2024,Green2025}, which serves as the required input for a quantum computation. This task is fundamental to quantum algorithm design, as the efficiency and accuracy of state preparation directly influence the overall performance and scalability of quantum computations \cite{zhang2022, OBrien2025quantumstate,Bausch2022fastblackboxquantum}.
Tensor networks (TNs) provide an efficient framework for representing complex quantum states in a compact classical form, making them a natural foundation for designing shallow-depth quantum circuits capable of preparing complex and entangled states. This connection has motivated extensive research into the preparation of various TN architectures, with particular emphasis on Matrix Product States (MPS) \cite{schon2005,Rudolph2022,BenDov2024,iaconis2024,Malz2024,Smith2024,melnikov2023,Jaderberg2025,Jobst2024efficientmps,Green2025}, as well as Tree Tensor Networks (TTN) \cite{Sugawara2025,Manabe2025,Schuhmacher_2025}, the Multi-Scale Entanglement Renormalization Ansatz (MERA) \cite{vidal2008,cincio2008}, and, more recently, two-dimensional isometric Projected Entangled Pair States (isoPEPS) \cite{wei_malz2022,Schwarz2012,yu2024}. 

The MPS is one of the most widely studied of all TNs, providing an efficient representation of certain classes of one-dimensional area-law quantum states, such as the ground-states of gapped local Hamiltonians \cite{Cirac2021,verstraete2006,Eisert2010AreaLaws,yu2017}. The MPS possesses many advantageous computational features, such as the ability to efficiently compute canonical forms and Schmidt decompositions across its bipartitions. Despite being formally tailored to one-dimensional systems, these powerful computational features often make MPS the leading ansatz in the numerical study of certain two-dimensional quantum systems, such as the 2D Hubbard or Heisenberg models \cite{stoudenmire2012,ehlers2017,WhiteScalapino1998}. 

It is well-known that any one-dimensional isometric TN with bond dimension $\chi$ can be realised by a quantum circuit through the embedding of all tensors into $\mathcal{O}(\log_2\chi+1)$-qubit gates \cite{schon2005}. The topology and circuit depth of such a mapping typically matches the topology of the TN itself (i.e., linear-depth for MPS, logarithmic for TTN and MERA with each branch being prepared in parallel). However, the decomposition of multi-qubit gates into smaller, physically realisable gates can lead to significant computational overhead \cite{Mottonen2004,Krol2022,Ran2020}. For example, although exact MPS synthesis scales efficiently as $\mathcal{O}(n\chi^2)$, the requirement for decomposing multi-qubit gates typically leads to significantly deeper-than-necessary quantum circuits \cite{Green2025,Ran2020,BenDov2024}.

The first significant work-around to multi-qubit decompositions was the development of the Matrix Product Disentangler (MPD) algorithm, designed to approximately disentangle an MPS towards a product state using sequential layers of local one- and two-qubit gates \cite{Ran2020}. After the MPD is computed, the disentangling process can be reversed, generating a circuit that progressively reconstructs the entanglement of the target MPS. However, the MPD algorithm is pathological: its success depends on the assumption that the operator that disentangles a low-rank MPS approximation can also effectively disentangle the MPS it approximates, which is not generally true. Further, the MPD often suffers from an exponential scaling of computational resources with the number of circuit layers, largely owing to the suboptimality of the disentangling process \cite{Green2025,BenDov2024}.

We introduce Schmidt Spectrum Optimisation (SSO) as an improved framework for efficiently disentangling MPS using sequential circuit layers. In SSO, disentangling is reformulates as an optimisation problem defined explicitly on the Schmidt spectra, maximising the fidelity of the $\chi=2$ approximation to the intermediate disentangled MPS at each step. After complete disentangling, an efficient state preparation circuit can be easily obtained by reversing the learned disentangling circuit. A related approach, Classical Variational Disentanglement (CVD) \cite{mansuroglu2026}, instead uses the Rényi entropy of the intermediate state as its target. We discuss this method in Sec.~\ref{sec:related_work} and benchmark against it in Sec.~\ref{sec:results}, finding that the SSO approach exhibits superior performance across core benchmarks. 

We further find that the SSO algorithm computes significantly more effective disentangling layers when compared to the baseline MPD algorithm. Further, it is a known issue that disentangling methods suffer from worst-case exponential growth of bond dimension with circuit depth -- an artefact of repeatedly contracting intermediate MPS with disentangling operators. However, this issue is mitigated when such disentangling operators succeed in reducing entanglement in each MPS, allowing the intermediate bond dimension to be truncated significantly. Crucially, we find that the improved effectiveness of the SSO algorithm in computing disentangling layers significantly improves the scaling of bond dimension with circuit depth, allowing it to be scaled to deeper circuits efficiently. 

We evaluate SSO on a range of state preparation tasks, including random MPS, and MPS approximations to the ground-states of one-dimensional and two-dimensional local Hamiltonians on up to 60 qubits. Across these benchmarks, SSO achieves state-of-the-art shallow-depth performance relative to the MPD algorithm \cite{Ran2020} and the CVD algorithm \cite{mansuroglu2026}. We also show that the SSO algorithm outperforms optimised variants of the MPD algorithm \cite{Rudolph2022,BenDov2024}, and that a joint optimisation of circuit layers initialised by the output of the SSO algorithm further improves results. Hence, we propose the SSO algorithm as an improved and scalable approach to quantum state preparation.

\section{Matrix Product States}

We provide a minimal background on the form, Schmidt decomposition, and bond dimension truncation of matrix product states. For a more rigorous theoretical background, we point the reader to \cite{orus2014,Oseledets2011}.

\textbf{\textit{Definition.}} Consider a one–dimensional quantum system of $n$ sites with local Hilbert space $\mathbb{C}^d$ and computational basis $\{\lvert i\rangle\}_{i=0}^{d-1}$. For qubits, $d=2$. A Matrix Product State (MPS) with open boundary conditions and virtual dimensions $\{\alpha_j\}_{j=0}^n$ (with $\alpha_0=\alpha_n=1$) is the variational ansatz
\begin{equation}
\label{eq:mps-def}
\lvert \psi \rangle
= \sum_{i_1,\dots,i_n=0}^{d-1} \;\sum_{\{\alpha_j\}}
A^{[1]\, i_1}_{\alpha_0,\alpha_1}
A^{[2]\, i_2}_{\alpha_1,\alpha_2}
\cdots
A^{[n]\, i_n}_{\alpha_{n-1},\alpha_n}\,
\lvert i_1 i_2 \cdots i_n \rangle,
\end{equation}
where for each site $j$ and physical index $i_j$, $A^{[j]\, i_j}_{\alpha_{j-1},\alpha_j}\in \mathbb{C}^{\alpha_{j-1} \times \alpha_j}$ is a complex matrix \cite{orus2014}. The bond dimension $\chi:=\arg\max_j \alpha_j$ is defined as the maximum virtual dimension in the MPS. The MPS contains $\mathcal{O}(2n\chi^2)$ parameters, and $\chi$ controls the expressivity of the MPS.

\textbf{\textit{Canonical form.}} An MPS is said to be in left-canonical form up to site $m$ if, for each $j\le m$,
\begin{equation}
\label{eq:left-canon}
\sum_{i=0}^{d-1} \bigl(A^{[j]\, i}\bigr)^\dagger A^{[j]\, i} = I_{\alpha_j}.
\end{equation}
The left-canonical constraint in \eqref{eq:left-canon} states that the linear map from the left bond space to the composite (right bond $\otimes$ physical) space is an isometry. Similarly, right-canonical tensors satisfy $\sum_i A^{[j]\, i}\bigl(A^{[j]\, i}\bigr)^\dagger = I_{\alpha_{j-1}}$. Any MPS with open-boundary conditions can be put into canonical form using successive QR or SVD factorisations with $\mathcal{O}(n\chi^3)$ complexity \cite{Schollwock2011,Cirac2021}.

\begin{figure}[b]
    \centering
    \includegraphics[width=1\linewidth]{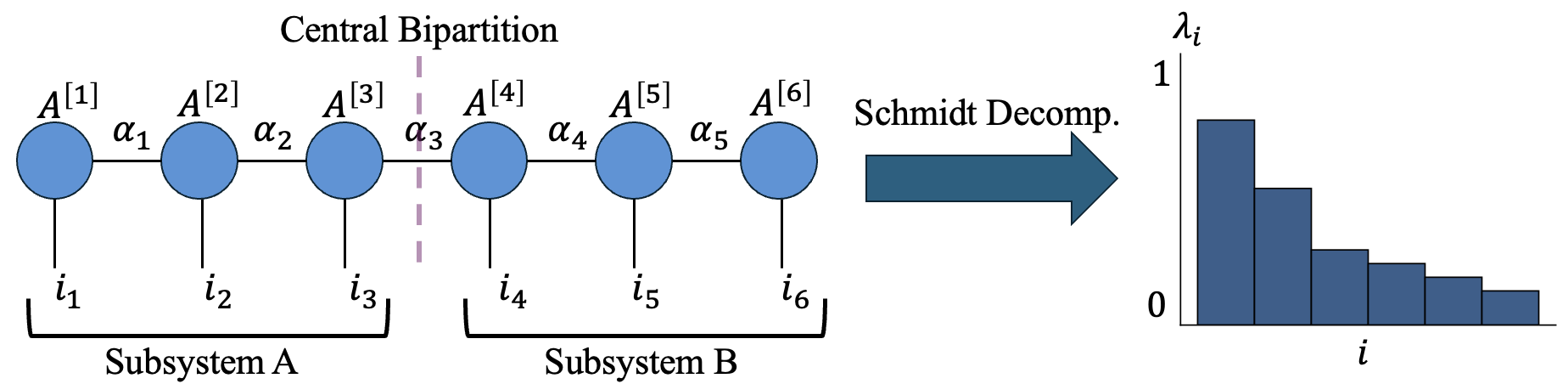}
    \caption{Computing the Schmidt decomposition for a specified bipartition of the MPS with virtual dimension $|\alpha_j|$ generates a vector of $|\alpha_j|$ non-increasing, non-negative Schmidt coefficients, which are used to define the loss function in the SSO algorithm. Here we visualise the Schmidt spectrum corresponding to the central bipartition (across $\alpha_3$) for a random real-valued MPS with $n=6$, $\chi=6$.}
    \label{fig:example_schmidt_decomposition}
\end{figure}

\textbf{\textit{Schmidt decomposition.}} 
Given a bipartition of the $n$-qubit system into subsystems $A$ and $B$ (for instance, $A = \{1,\dots,m\}$ and $B=\{m{+}1,\dots,n\}$), any pure state $\lvert \psi \rangle \in \mathcal{H}_A \otimes \mathcal{H}_B$ admits a Schmidt decomposition
\begin{equation}
\label{eq:schmidt}
\lvert \psi \rangle
= \sum_{k=1}^{r} \lambda_k\,
\lvert \Phi_k \rangle_A \otimes \lvert \Phi_k \rangle_B,
\end{equation}
where the Schmidt coefficients $\lambda_k \ge 0$ are non-increasing, satisfy $\sum_k \lambda_k^2 = 1$, and $r = \mathrm{rank}(\rho_A) = \mathrm{rank}(\rho_B)$ is the Schmidt rank. We denote the square of Schmidt coefficients, $p_k := \lambda_k^2$, as the Schmidt probabilities. The von Neumann entanglement entropy across the bipartition is
\begin{equation}
\label{eq:entropy}
S(A{:}B) \;=\; - \sum_k p_k \log p_k
\;\;\le\;\; \log r \;\le\; \log \alpha_m,
\end{equation}
where $\alpha_m$ is the bond dimension of the virtual index cut by the bipartition. The Schmidt spectrum $\{\lambda_k\}_k$ completely characterises the entanglement across that cut \cite{Oseledets2011,Cirac2021}. An example Schmidt decomposition across the central bipartition of a 6-site MPS is visualised in Fig.~\ref{fig:example_schmidt_decomposition}.

\textbf{\textit{Computing the Schmidt decomposition.}} 
For an MPS, the bipartition $A|B$ induced by cutting a single virtual bond can be put in \emph{mixed canonical form}. In practice, one left-orthonormalises tensors on $A$ and right-orthonormalises tensors on $B$ via successive QR or SVD steps. The tensor at the cut is then reshaped into a matrix and decomposed with a single SVD; its singular values are exactly the Schmidt coefficients for that bipartition. Further details can be found in \cite{orus2014,Oseledets2011,Schollwock2011}.

\textbf{\textit{Bond-dimension truncation.}} 
Consider truncating the virtual bond associated with $A|B$ from bond dimension $\alpha_m$ to $\alpha_m' < \alpha_m$. The Frobenius-norm optimal approximation is obtained by retaining only the $\alpha_m'$ largest Schmidt coefficients \cite{Mirsky1960,EckartYoung1936}. If we order the coefficients such that $\lambda_1 \ge \lambda_2 \ge \dots \ge \lambda_{\alpha_m}$, the truncation error is 
\begin{equation}
\label{eq:truncation_error}
\varepsilon_{\mathrm{disc}}(\alpha_m') \;:=\; \sum_{k>\alpha_m'} p_k
\;=\; \sum_{k>\alpha_m'} \lambda_k^2,
\end{equation}
which equals the squared norm of the discarded component. The fidelity between the state $\lvert \psi \rangle$ and the (renormalised) truncated state $\lvert \psi_{\chi'} \rangle$ is then $F(\alpha_m') = 1 - \varepsilon_{\mathrm{disc}}(\alpha_m')$.

\section{Relation to Previous Work}
\label{sec:related_work}
We now contextualise our work and contributions in relation to previous MPS-based state preparation approaches to clarify the scope, novelty, and core advantages of the Schmidt Spectrum Optimisation (SSO) algorithm.

\subsection{Exact Synthesis and Parallelisation of MPS}

It is well known that any MPS of bond dimension $\chi$ can be prepared exactly by a single sequential (staircase) circuit of local $\mathcal{O}(\log_2 \chi + 1)$-multi-qubit gates \cite{schon2005}. More recently, Malz \emph{et al.} \cite{Malz2024} have shown that this linear-depth scaling is not fundamental: for translation-invariant normal MPS, they prove a lower bound $T = \Omega(\log N)$ on the circuit depth of any local-unitary preparation protocol and construct a renormalisation-group-inspired algorithm that prepares such states with error $\epsilon$ in depth $T = \mathcal{O}[\log(N/\epsilon)]$, which is optimal. Their scheme can be further accelerated using mid-circuit measurements and feedback, achieving depth $T = \mathcal{O}[\log\log(N/\epsilon)]$ in this adaptive setting and extending to a broader class of MPS (including some non-normal and inhomogeneous instances).

Complementary to these log-depth constructions, Smith \emph{et al.} \cite{Smith2024} show that adaptive circuits combining local unitaries, mid-circuit measurements and feedforward can exactly prepare a broad class of MPS in \emph{constant} depth, including short- and long-range entangled phases, symmetry-protected/topological states and MBQC resource states. Exploiting global on-site symmetries, they characterise when constant-depth preparation is possible and give explicit adaptive protocols that outperform any unitary-only scheme. Taken together, these works establish tight asymptotic depth bounds for exact preparation of structured MPS families. By contrast, here we address the \emph{approximate}, shallow-depth preparation of generic target MPS using fixed two-qubit gate layers, where constant or logarithmic depth is typically unattainable and the central challenge is to obtain numerically stable, high-fidelity circuits under strict depth constraints.

\subsection{The Matrix Product Disentangler (MPD) Algorithm}

Decomposing the multi-qubit gates in the exact MPS synthesis circuit into one- and two-qubit gates yields a circuit of depth $\mathcal{O}(n \chi^2)$. While efficient in complexity terms, this decomposition introduces a large constant overhead in practice \cite{Krol2022}. This motivated the so-called Matrix Product Disentangler (MPD) algorithm, which bypasses transpilation by directly constructing a state-preparation circuit of one- and two-qubit gates at the expense of an approximation error \cite{Ran2020}. In many cases, the MPD algorithm yields shallower depth circuits when compared to the exact decomposition approach \cite{Green2025}.

In the MPD algorithm, the intermediate MPS $\ket{\psi^{(k)}}$ at disentangling layer $k$ is compressed to its optimal $\chi=2$ approximation $\ket{\tilde{\psi}^{(k)}}$. The $k^{\text{th}}$ disentangling operator is then defined such that
\begin{equation}
U_k\ket{\tilde{\psi}^{(k)}}=\ket{0}^{\otimes n}
\end{equation}
where $U_k$ is implemented as a single layer of local $SU(4)$ gates. Since $\ket{\tilde{\psi}^{(k)}}$ \textit{approximately} mirrors the entanglement structure of  $\ket{\psi^{(k)}}$, $U_k$ is expected to roughly disentangle $\ket{\psi^{(k)}}$. The intermediate state is then updated as
\begin{equation}
    \ket{\psi^{(k+1)}}\leftarrow U_k\ket{\psi^{(k)}}
\end{equation}
and the disentangling process is repeated for $k=1,\dots,L$. The adjoint circuit acts such that
\begin{equation}
    U_1^\dagger U_2^\dagger \cdots U_L^\dagger\ket{0}^{\otimes n}\approx\ket{\psi^{(1)}}
\end{equation}
where $\ket{\psi^{(1)}}$ is the target MPS. We define the accuracy of this state preparation protocol using the fidelity $F:=|\braket{\psi^{(1)}|U_1^\dagger U_2^\dagger \cdots U_L^\dagger|0}|^2$. This general approach to state preparation based on initially disentangling a target state is likewise the foundational idea behind the proposed SSO algorithm.

The MPD algorithm suffers from two key pathologies. Firstly, $U_k$ effectively disentangles $\ket{\psi^{(k)}}$ when the fidelity
\begin{equation}
F_{\chi=2}=|\braket{\psi^{(k)}|\tilde{\psi}^{(k)}}|^2
\end{equation}
is close to 1. As $F_{\chi=2}\rightarrow0$, the entanglement structure of $\ket{\tilde{\psi}^{(k)}}$ ceases to be representative of $\ket{\psi^{(k)}}$, and $U_k$ becomes an ineffective disentangler. For many physically relevant target states of interest, $F_{\chi=2}$ is small, so the MPD produces ineffective disentangling layers. Secondly, if $\ket{\psi^{(k)}}$ has bond dimension $\chi^{(k)}$, then applying $U_k$ yields a state $\ket{\psi^{(k+1)}}$ with bond dimension $\chi^{(k+1)}=2\chi^{(k)}$. When the disentangling layer is effective at reducing entanglement in the target state, $\chi^{(k+1)}$ can often be truncated back down without loss. In practice, however, the MPD algorithm often produces suboptimal disentangling layers, leading to exponential growth of the bond dimension with the number of layers $L$. In Sec.~\ref{sec:results}, we show that the SSO algorithm significantly mitigates these pathologies.

\subsection{Tensor Network Optimisation (TNO) Approaches}

More recent work has proposed preparing MPS using fully classical tensor network optimisation (TNO) frameworks, in which the parameterised quantum circuit is represented explicitly as a tensor network \cite{melnikov2023,BenDov2024,Rudolph2022}. Since the target state is an MPS, the fidelity
\begin{equation}
F_S=|\braket{\psi_{\text{targ}}|U(\theta)|0^{\otimes n}}|^2
\end{equation}
can be evaluated entirely classically. However, many TNO approaches are known to suffer from optimisation pathologies, including severe local minima \cite{Green2025} and loss landscapes that exhibit features analogous to barren plateaus \cite{liu2022}. Fortunately, the MPD algorithm can be used to initialise TNO parameters $\theta_0=\theta_{\text{MPD}}$, which has been shown to significantly improve optimisation outcomes \cite{BenDov2024,Rudolph2022,Green2025}. We consider these optimised variants of the MPD algorithm when benchmarking SSO in Sec.~\ref{sec:results}.

\subsection{The Classical Variational Disentanglement (CVD) Algorithm}

The Classical Variational Disentanglement (CVD) algorithm \cite{mansuroglu2026} is an independent and concurrent work that follows the same core idea as SSO: using classically efficient optimisation to learn sequential operators that remove entanglement from the target state. While sharing the same foundational idea of variational preparation-by-disentangling, the SSO and CVD algorithms differ in two crucial ways. 

Firstly, the CVD algorithm uses a brick-wall ansatz, as opposed to the staircase-like ansatz adopted in our work. More critically, the two approaches differ in objective: the CVD algorithm introduces a cost function based on minimising the Rényi Entropy of intermediate states. In our work, we introduce a novel objective based on minimising the error of the $\chi=2$ approximation to the disentangled state. The core motivation for adopting such an objective is that we can leverage the analytical mapping from any $\chi=2$ MPS to the all-zero computational state with a single layer of 2-qubit gates \cite{schon2005}. This layer of 2-qubit gates takes the form of a staircase-like ansatz, and therefore our choice of ansatz is in harmony with the chosen objective. We suggest that the objective adopted by the SSO algorithm is fundamentally easier from an optimisation standpoint, and the $\chi=2$ approximation to the final disentangled MPS being analytically mapped to a product state provides accuracy guarantees that cannot be ensured by the CVD algorithm. Specifically, the SSO objective is a tight upper bound on the infidelity of the prepared state (Theorem~\ref{thm:fidelity_guarantee}), whereas the Rényi-entropy cost design of CVD admits no comparable bound. We substantiate these claims empirically through benchmarking in Sec.~\ref{sec:results}.

\section{Schmidt Spectrum Optimisation (SSO)}

We now present a detailed description of the Schmidt Spectrum Optimisation (SSO) algorithm. We describe the structure of SSO, examine its cost function, explore post-processing strategies, and discuss its computational complexity, situating the method within the broader landscape of tensor-network optimisation.

\begin{figure*}[t]
    \centering
    \includegraphics[width=1\linewidth]{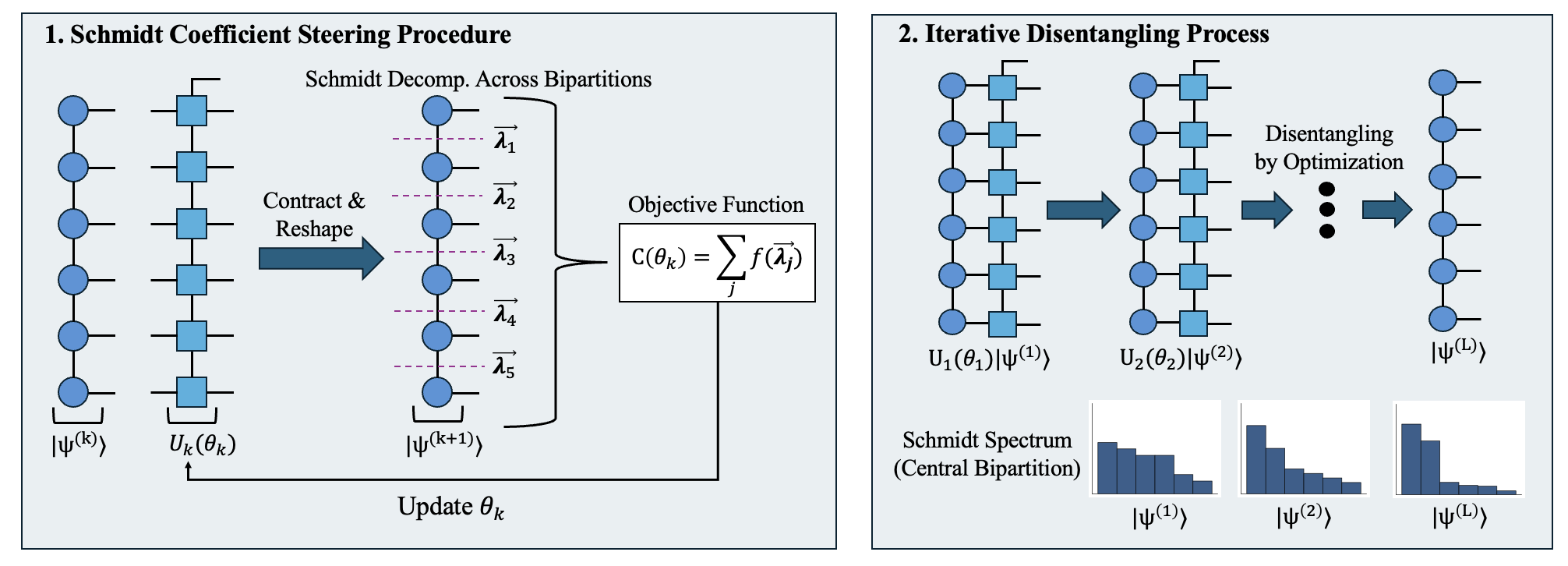}
    \caption{The \emph{Schmidt Spectrum Optimisation} (SSO) algorithm. The parameterized circuit layer $U_k(\theta_k)$ is contracted with the intermediate MPS $\ket{\psi^{(k)}}$ producing an updated MPS $\ket{\psi^{(k+1)}}$. The Schmidt decomposition is computed across all bipartitions of $\ket{\psi^{(k+1)}}$, and the circuit parameters $\theta_k$ are updated via gradient descent to minimise the specified objective function. The specific objective function we introduce in Eq.~\ref{eq:new_cost_fn} corresponds to $f(\vec{\lambda}_j)=\vec{\lambda}_{1,j}^2+\vec{\lambda}_{2,j}^2$. By repeating this process, the MPS is iteratively disentangled.}
    \label{fig:overarching_figure}
\end{figure*}

\subsection{Disentangling via Schmidt Spectrum Optimisation}

Let $\ket{\psi^{(k)}}$ be an $n$-qubit quantum state represented as an MPS with bond dimension $\chi^{(k)}$. For each bipartition between sites $i$ and $i+1$ there is a Schmidt decomposition with Schmidt values $\{\lambda_{i,j}^{(k)}\}$, which we collect into a vector $\vec{\lambda_i}^{(k)}$ ordered such that $\lambda_{i,1}^{(k)}\geq\lambda_{i,2}^{(k)}\geq\cdots\geq\lambda_{i,|\alpha_i|}^{(k)}$ and $\sum_{j=1}^{|\alpha_i|}\left(\lambda_{i,j}^{(k)}\right)^2=1$. We denote the full set of Schmidt spectra across all $n-1$ possible bipartitions of $\ket{\psi^{(k)}}$ by $\boldsymbol{\lambda}^{(k)}=\{\vec{\lambda_i}^{(k)}\}_{i=1}^{n-1}$. 

Starting with a target MPS $\ket{\psi^{(1)}}$, the objective of SSO is to optimise $L$ layers of local unitaries $\{U_k(\theta_k)\}_{k=1}^{L}$ that sequentially remove entanglement from the state. Specifically, the objective at iteration $k$ is to optimise $\theta_k$ such that the updated state
\begin{equation}
    \ket{\psi^{(k+1)}}= U_k(\theta_k)\ket{\psi^{(k)}}
\end{equation}
has Schmidt spectra $\boldsymbol{\lambda}^{(k+1)}$ that extremise a cost function $C(\boldsymbol{\lambda}^{(k+1)})$.

We adopt the (staircase) circuit ansatz $U_k(\theta_k)$ consisting of a single layer of parameterized $SU(4)$ gates, which is the $\chi=2$ circuit subclass of Matrix Product Unitary (MPU) \cite{Styliaris2025}:
\begin{equation}
    U_k(\theta_k) =\prod_{i=1}^{n-1}\Big(\mathbb{I}^{\otimes (i-1)}\otimes
U^{[i,i+1]}(\theta_{k,i})\otimes\mathbb{I}^{\otimes (n-i-1)}\Big)
\label{eq:staircase_ansatz}
\end{equation}
where each 2-qubit gate $U^{[i,i+1]}(\boldsymbol{\theta}_{k,i})$ requires 15 parameters to fully parameterize.

To evaluate the cost $C(\boldsymbol{\lambda}^{(k+1)})$ and its gradients, we need access to the Schmidt spectra of $\ket{\psi^{(k+1)}}$. The staircase ansatz in Eq.~\ref{eq:staircase_ansatz} allows each two-qubit gate $U^{[j,j+1]}(\theta_{k,j})$ to be locally contracted with the neighbouring MPS tensors with physical indices $i_j$ and $i_{j+1}$. This keeps $\ket{\psi^{(k+1)}}$ in MPS form using only local contractions, so $\ket{\psi^{(k+1)}}$ can be brought into mixed canonical form and the Schmidt values $\{\lambda_{i,j}^{(k+1)}\}$ can be computed. Automatic differentiation through these tensor contractions yields stable gradients $\nabla_{\theta_k}C(\boldsymbol{\lambda}^{(k+1)})$, and the circuit parameters $\theta_k$ can be trained via gradient descent \cite{pytorch}. This methodology is summarised in Fig.~\ref{fig:overarching_figure}.

\subsection{Tail Loss Function and Optimisation}

\begin{figure}[h!]
    \centering
    \includegraphics[width=1\linewidth]{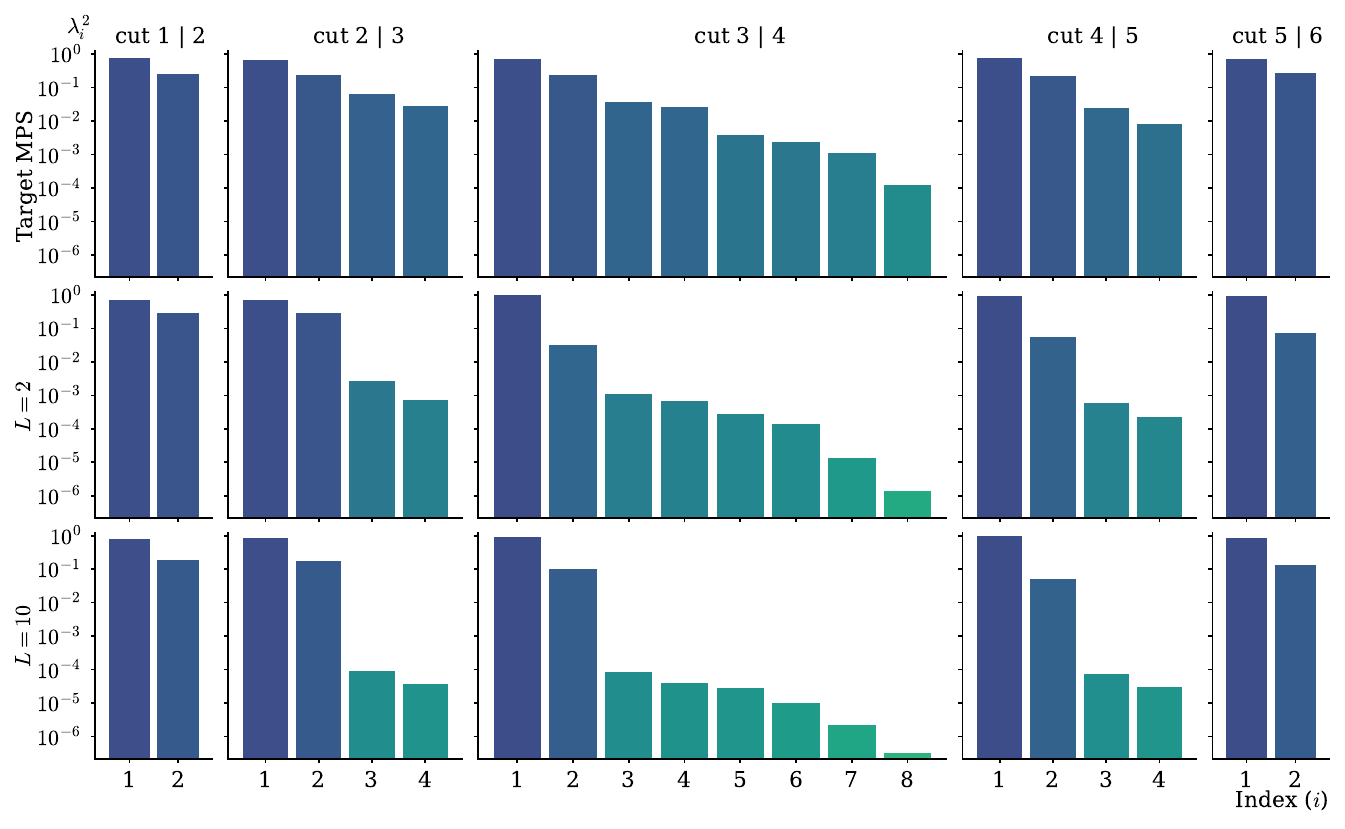}
    \caption{The normalised Schmidt probabilities ($\lambda_i^2$) across each bipartition of a 6-site random MPS and its disentangled forms, computed via the SSO algorithm and the loss function defined in Eq.~\ref{eq:new_cost_fn}. In this example, the target MPS has $F_{\chi=2}=0.8356$ and the disentangled state after $L=10$ layers has $F_{\chi=2}=0.9998$.}
    \label{fig:schmidt_evolution}
\end{figure}

We choose a loss function that encourages each bipartition of $\ket{\psi^{(k+1)}}$ to be well approximated by a rank-2 truncation. The optimal $\chi=2$ truncation error at bond $i$ is given by
\begin{equation}
\epsilon_i^{(k+1)}=\sum_{j>2}\left(\lambda_{i,j}^{(k+1)}\right)^2=1-\left(\lambda_{i,1}^{(k+1)}\right)^2-\left(\lambda_{i,2}^{(k+1)}\right)^2
\end{equation}
where $\lambda_{i,1}^{(k+1)}$ and $\lambda_{i,2}^{(k+1)}$ are the largest and second largest coefficients of the Schmidt decomposition between sites $i$ and $i+1$, respectively. We define the layer-$k$ loss as the sum of these local truncation errors over all bonds:
\begin{equation}
C(\boldsymbol{\lambda}^{(k+1)})=\sum_{i=1}^{n-1}1-\left(\lambda_{i,1}^{(k+1)}\right)^2-\left(\lambda_{i,2}^{(k+1)}\right)^2
    \label{eq:new_cost_fn}
\end{equation}
This loss is implicitly parameterized by $\theta_k$. Minimising the loss in Eq.~\ref{eq:new_cost_fn} amounts to maximising, at each bond, the weight retained by the leading two Schmidt coefficients, and hence minimising the local truncation error of the $\chi=2$ approximation to $\ket{\psi^{(k+1)}}$ (cf. Theorem~\ref{thm:fidelity_guarantee}). See a visualisation of the concentration of the Schmidt weight in the leading two coefficients for a 6-site random MPS in Fig.~\ref{fig:schmidt_evolution}.

We initialise each circuit layer by sampling $\theta_k\sim\mathcal{N}(0,10^{-3})$ such that $U_k(\theta_k)\approx\mathbb{I}_{2^n}$ and $\boldsymbol{\lambda}^{(k+1)}\approx\boldsymbol{\lambda}^{(k)}$ at initialisation. This prevents the initial increase in entanglement that would be induced by a random parameter initialisation. We learn parameters using the L-BFGS-B optimiser. We also experiment with an alternative loss function based on a Rényi entropy target, though we show in Fig.~\ref{fig:comparison_with_cvd}(a--b) that the tail loss function exhibits superior performance at fixed circuit depth. Notably, this control suggests that the primary performance difference between the SSO and CVD algorithms is not the objective function alone. Rather, the advantage stems from pairing the tail-loss objective with the staircase ansatz, whose final layer maps the $\chi=2$ approximation of the disentangled state analytically to a product state, fixing the prepared fidelity at exactly that approximation (cf. Eq.~\ref{eq:fidelity_equality}) -- a guarantee the CVD algorithm does not provide.

\subsection{State Preparation by Reversing Disentangling}

\begin{figure}[!]
    \centering
    \includegraphics[width=1\linewidth]{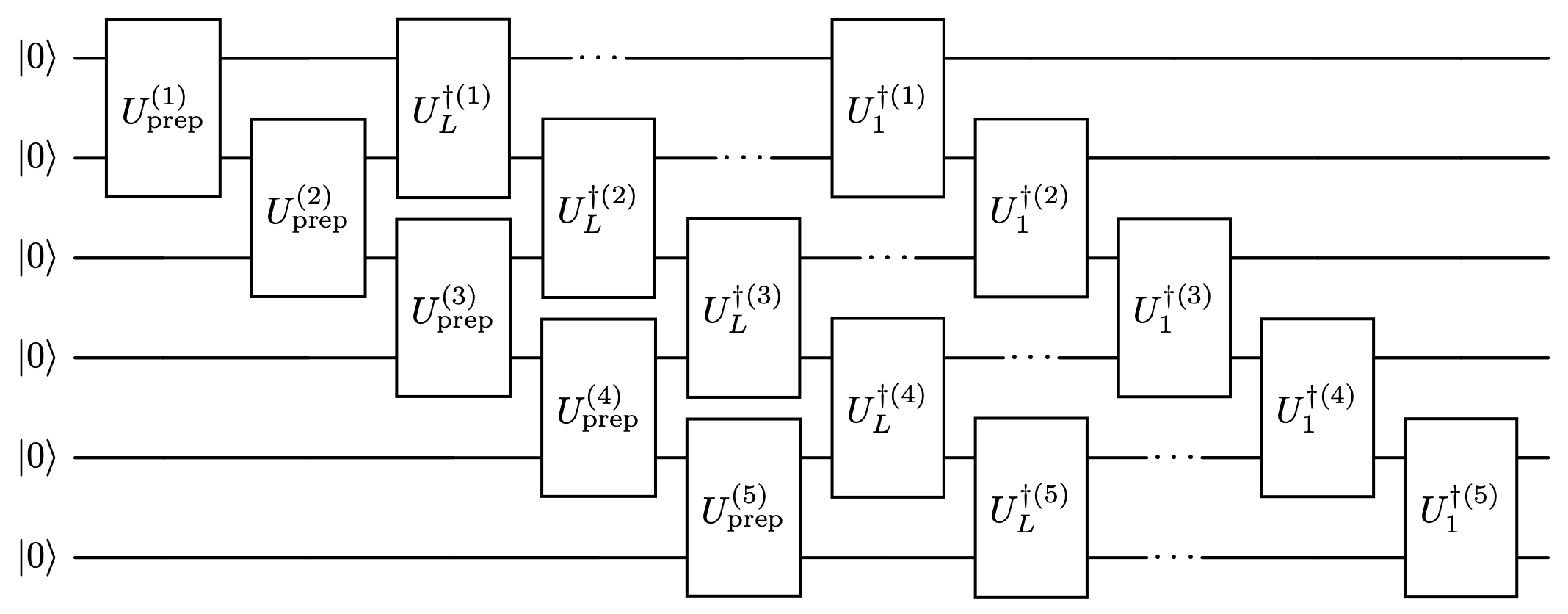}
    \caption{The quantum circuit generated by the SSO algorithm for $L$ optimised layers and 6 qubits, showing how each staircase-like layer can be parallelised. The output of the circuit is an approximation of the target MPS.}
    \label{fig:circuit_diagram}
\end{figure}

The objective of the SSO algorithm is to compute a set of $L$ efficient unitary operators $\{U_k\}_{k=1}^L$ that sequentially disentangle the target MPS $\ket{\psi^{(1)}}$. After $L$ iterations we obtain
\begin{equation}
    \ket{\psi^{(L)}}=U_L \cdots U_2 U_1\ket{\psi^{(1)}}
\end{equation}
where $\ket{\psi^{(L)}}$ is designed to be well approximated by a $\chi=2$ MPS $\ket{\tilde{\psi}^{(L)}}$. The corresponding fidelity of this low-rank approximation is 
\begin{equation}
F_L:=|\braket{\psi^{(L)}|\tilde{\psi}^{(L)}}|^2
\end{equation}
To reverse the disentangling process, we first prepare $\ket{\tilde{\psi}^{(L)}}$ using
\begin{equation}
    U_{\text{prep}}\ket{0}^{\otimes n}=\ket{\tilde{\psi}^{(L)}}
\end{equation}
where $U_{\text{prep}}$ can be computed exactly by embedding the tensors of $\ket{\tilde{\psi}^{(L)}}$ into a single layer of $SU(4)$ gates (i.e., the exact same circuit form of a single layer of the staircase ansatz in Eq.~\ref{eq:staircase_ansatz}). Then, the full state preparation circuit is given by
\begin{equation}
    U_S:=U_1^\dagger U_2^\dagger \cdots U_L^\dagger U_{\text{prep}}
\end{equation}
such that $U_S\ket{0}^{\otimes n}$ prepares an approximation of the original target $\ket{\psi^{(1)}}$. See Fig.~\ref{fig:circuit_diagram} for a visualisation, noting that the sequential circuit structure allows for operations to be run in parallel across all qubits. This is directly analogous to the MPD algorithm \cite{Ran2020} but with optimised disentangling operators. By unitarity, the fidelity of the approximate state prepared by $U_S$ is exactly
\begin{equation}
|\braket{\psi^{(1)}|U_S|0^{\otimes n}}|^2=|\braket{\psi^{(L)}|\tilde{\psi}^{(L)}}|^2=F_L
\label{eq:fidelity_equality}
\end{equation}
Thus, minimising the loss in Eq.~\ref{eq:new_cost_fn} directly maximises the fidelity of the final prepared state with the target, subject to the shallow-depth constraint. Note that a target circuit depth of $L$ layers includes only $L-1$ optimised layers, followed by the additional layer $U_{\text{prep}}$ preparing $\ket{\tilde{\psi}^{(L)}}$.

\begin{theorem}[Fidelity guarantee]
\label{thm:fidelity_guarantee}
Let $C^{(L)} = \sum_{i=1}^{n-1} \epsilon_i^{(L)}$ denote the final value of the
tail loss in Eq.~\ref{eq:new_cost_fn}, where
$\epsilon_i^{(L)} = 1 - \bigl(\lambda_{i,1}^{(L)}\bigr)^{2} - \bigl(\lambda_{i,2}^{(L)}\bigr)^{2}$.
The fidelity of the SSO state preparation circuit satisfies
\begin{equation}
F_S \;\geq\; \bigl(1 - C^{(L)}\bigr)^{2} \;\geq\; 1 - 2\,C^{(L)}.
\label{eq:fidelity_bound}
\end{equation}
\end{theorem}

\begin{proof}
By Eq.~\ref{eq:fidelity_equality}, $F_S = \bigl|\langle \psi^{(L)} \mid \tilde{\psi}^{(L)} \rangle\bigr|^{2}$,
where $|\tilde{\psi}^{(L)}\rangle$ is obtained from $|\psi^{(L)}\rangle$ by truncating
each virtual bond to dimension $\chi = 2$. The sequential-SVD truncation bound gives
\begin{equation}
\bigl\| \, |\psi^{(L)}\rangle - |\tilde{\psi}^{(L)}\rangle \, \bigr\|^{2}
\;\leq\; 2 \sum_{i=1}^{n-1} \epsilon_i^{(L)} \;=\; 2\, C^{(L)},
\end{equation}
hence $\langle \psi^{(L)} \mid \tilde{\psi}^{(L)} \rangle \geq 1 - C^{(L)}$.
Squaring yields the stated bound.
\end{proof}

\subsection{Fidelity-Maximising Post-Processing (SSO+All)}

In the SSO algorithm, each layer is optimised in isolation while keeping the parameters of all other layers fixed. SSO is therefore a greedy algorithm in the sense that each layer minimises the loss in Eq.~\ref{eq:new_cost_fn} without collaboration with other layers. As such, it is possible to improve the fidelity of the prepared state by jointly optimising all layers as a post-processing step. The fidelity-maximising loss function is defined as
\begin{equation}
F(\theta_{1:L})=|
\braket{\psi^{(1)}|U_S(\theta_{1:L})|0^{\otimes n}}|^2
\label{eq:fidelity_postprocessing}
\end{equation}
where $U_S(\theta_{1:L})$ represents the circuit consisting of all $L$ parameterized layers (including $U_{\text{prep}}$). The parameters $\theta_{1:L}$ are initialised to be exactly those computed by the SSO algorithm. We denote this algorithm as SSO+All, which we benchmark in Sec.~\ref{sec:results}.

Unless efficient bond dimension truncations can be made in the contraction of the $L$ circuit layers, the bond dimension required to compute the loss in Eq.~\ref{eq:fidelity_postprocessing} scales as $\chi_{\text{max}}=\mathcal{O}(2^L)$. Moreover, collective optimisation of deep tensor network circuits is known to suffer from severe local minima problems and loss landscapes mirroring barren plateaus as depth is increased \cite{liu2022,Green2025,Green2025}. These effects are, however, diminished when the training is initialised by high-quality parameters \cite{Rudolph2022,Grant2019}. This implies that SSO+All can be efficiently trained for small numbers of layers $L$, and is particularly well-suited to near-term quantum implementation.

\subsection{Complexity Analysis}

\subsubsection{Time Complexity}
For an $n$-qubit target MPS $\ket{\psi^{(1)}}$ disentangled by $L$ layers with $T$ optimisation iterations per layer, the overall complexity of the SSO algorithm is
\[
\mathcal{O}\!\bigl(T\,n^2 L\,\chi_{\max}^3\bigr)
\]    
where $\chi_{\max}$ denotes the maximum bond dimension reached by any intermediate state during the disentangling process. The cubic dependence on $\chi_{\max}$ arises from the $\mathcal{O}(n \chi_{\max}^3)$ cost of computing Schmidt decompositions (or, equivalently, performing local SVDs) across the $n-1$ bonds of the MPS at each optimisation step.

Each circuit layer $U_k(\theta_k)$ can be viewed as the $\chi = 2$ instance of an MPU \cite{Styliaris2025}. When such a layer is contracted with an intermediate MPS $\ket{\psi^{(k)}}$ of bond dimension $\chi^{(k)}$, the resulting state $\ket{\psi^{(k+1)}}$ has bond dimension $\chi^{(k+1)} = 2 \chi^{(k)}$ before any truncation. Naïvely, this implies that the maximum bond dimension scales as $\chi_{\max} = \mathcal{O}(2^L \chi_{\text{targ}})$, so that the worst-case time complexity of the disentangling procedure would scale as $\mathcal{O}(8^L)$ in $L$. In SSO, however, each layer $U_k({\theta}_k)$ is explicitly optimised to \emph{disentangle} $\ket{\psi^{(k)}}$, so that in practice $\ket{\psi^{(k+1)}}$ can be truncated back to a much smaller bond dimension with negligible error. Hence, the scaling of $\chi_{\text{targ}}$ must be investigated empirically.

\subsubsection{Maximum Bond Dimension Scaling} As we show in Sec.~\ref{sec:results}, the extent to which the bond dimension can be truncated after each layer is directly tied to how effectively $U_k(\theta_k)$ reduces entanglement at layer $k$. To investigate this, we employ a simple truncation rule in which all Schmidt values satisfying $\lambda_{i,j}^{(k+1)} \le \lambda_{\text{thresh}}$ are discarded. In our experiments we set $\lambda_{\text{thresh}} = 10^{-7}$, ensuring that the truncation error is extremely small. For all SSO test cases considered, we observe that $\chi_{\max} = \mathcal{O}(\chi_{\text{targ}})$, i.e., the exponential growth in bond dimension with $L$ is completely avoided. This stands in sharp contrast to the MPD algorithm, where ineffective disentangling can lead to inefficient scaling of $\chi_{\text{max}}$, a claim which we substantiate in Sec.~\ref{sec:results}.

More aggressive truncation strategies can, in principle, further reduce the cost of SSO. For example, choosing a larger threshold $\lambda_{\text{thresh}}$ reduces $\chi_{\max}$ at the expense of increasing the truncation error (cf.\ Eq.~\ref{eq:truncation_error}) and thereby introducing an irreducible approximation error into the disentangling process. The optimal choice of $\lambda_{\text{thresh}}$ is problem-dependent, and in practice one may prefer adaptive schemes that adjust the allowed bond dimension dynamically based on local truncation errors or target accuracies. From a numerical perspective, the explicit control over Schmidt spectra and truncation thresholds also contributes to stable optimisation, preventing the ill-conditioned growth in bond dimension that can plague fidelity-maximising TNO approaches.

It is useful to contrast this behaviour with TNO schemes that treat the full $L$-layered circuit
\(
U(\theta_{1:L}) = U_L(\theta_L)\cdots U_1(\theta_1)
\)
as a single variational object and optimise all layers jointly to maximise the fidelity
\(
F(\theta_{1:L}) = \bigl|\langle \psi_{\text{targ}} | U(\theta_{1:L}) |0^{\otimes n}\rangle\bigr|^2.
\) The composed circuit $U(\theta_{1:L})$ is itself an isometric tensor network whose bond dimension grows as $\chi_{\text{circ}} = 2^L$ in the worst-case. Evaluating $F(\theta_{1:L})$ and its gradients exactly then requires contracting an MPS of bond dimension $\chi_{\text{circ}}$ with the target MPS, leading to a per-iteration cost that scales as $\mathcal{O}\bigl(n\,\chi_{\text{circ}}^3\bigr) = \mathcal{O}\bigl(n\,8^L\bigr)$. In practice, one must introduce truncations after intermediate layers to keep $\chi_{\text{circ}}$ manageable, but this couples approximation error directly to the optimisation dynamics. Hence, joint optimisation of all layers as a post-processing step is best-suited to shallow-depth circuits.

\section{Results}
\label{sec:results}

\begin{table*}[t]
    \centering
    \begin{tabular}{lccccc}
        \toprule
        Method      & Loss Function                 & Opt-Layerwise & Opt-All & Parameter Init. & Complexity* \\
        \midrule
        MPD \cite{Ran2020}        & --               & --                 & --            & None & $\mathcal{O}(nL\chi_{\text{max}}^3)$ \\
        CVD \cite{mansuroglu2026} & Rényi-$\alpha$ Entropy  & \checkmark & -- & Identity & $\mathcal{O}(Tn^2L\chi_{\text{max}}^3)$ \\
        SSO (\textit{Our Work})      & Schmidt Tail Loss             & \checkmark                 & --            & Identity & $\mathcal{O}(Tn^2L\chi_{\text{max}}^3)$\\
        MPD+LW \cite{Rudolph2022}      & Fidelity                      & \checkmark         & --            & MPD & $\mathcal{O}(TnL\chi_{\text{max}}^3)$\\
        MPD+All \cite{Rudolph2022,BenDov2024,Green2025}   & Fidelity                      & \checkmark         & \checkmark    & MPD+LW & $\mathcal{O}(T^2nL\chi_{\text{max}}^3)$\\
        SSO+All     & Schmidt + Fidelity    & \checkmark                 & \checkmark    & SSO & $\mathcal{O}(T^2n^2L\chi_{\text{max}}^3)$\\
        \bottomrule
    \end{tabular}
    \caption{Definitions of Compared Methods. \\ *$T$ denotes training iterations, $n$ qubits, $L$ layers, and $\chi_{\text{max}}$ maximum MPS bond dimension. Note that $\chi_{\text{max}}$ is a crucial parameter in the complexity, and the scaling of $\chi_{\text{max}}$ differs significantly depending on the state preparation methodology. For example, we show that SSO leads to a significantly reduced $\chi_{\text{max}}$ relative to MPD-based methods.}
    \label{table:method_definitions}
\end{table*}

In this section, we empirically assess the performance of Schmidt Spectrum Optimisation (SSO) against the Classical Variational Disentanglement (CVD), Matrix Product Disentangler (MPD) \cite{Ran2020} and optimised variants of the MPD algorithm \cite{Rudolph2022,BenDov2024}. The optimised MPD variants differ by optimisation approach. We denote the individual optimisation of MPD layers as MPD+LW, in which each circuit layer, initialised by the MPD algorithm, are optimised in isolation to maximise fidelity. We also consider a joint optimisation of all layers initialised by the MPD algorithm, which we denote by MPD+All. Likewise, we benchmark SSO+All, which first computes layers using the SSO algorithm before jointly optimising all layers further to maximise the fidelity of the prepared state. A summary of all methods is provided in Table~\ref{table:method_definitions}.

\subsection{Comparison with Classical Variational Disentanglement (CVD)}
\label{subsec:CVD}

\begin{figure*}[t]
    \centering
    \begin{subfigure}[t]{0.32\textwidth}
        \centering
        \includegraphics[width=\textwidth]{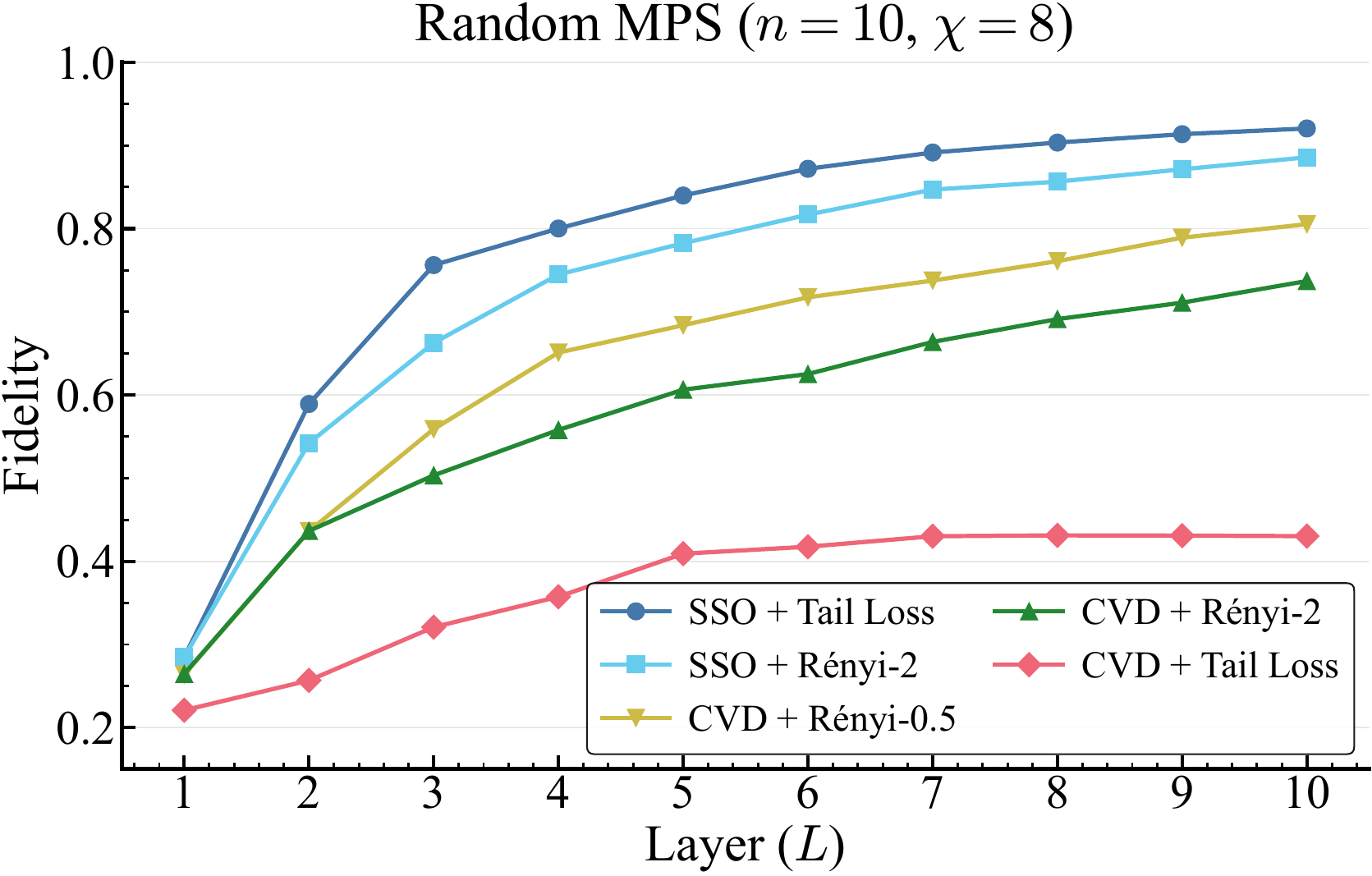}
        \caption{}
        \label{fig:sub1}
    \end{subfigure}
    \hfill
    \begin{subfigure}[t]{0.32\textwidth}
        \centering
        \includegraphics[width=\textwidth]{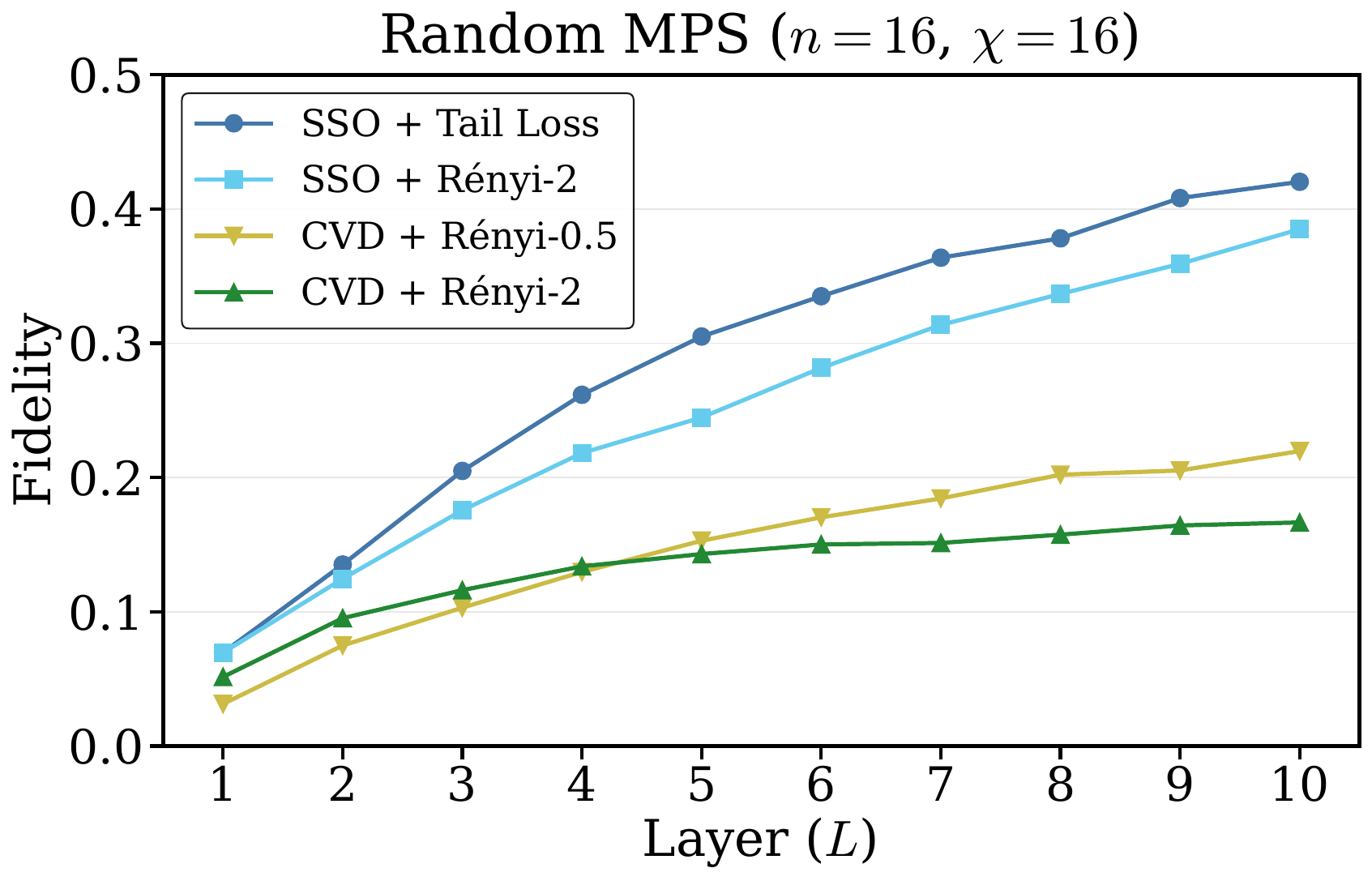}
        \caption{}
        \label{fig:sub2}
    \end{subfigure}
    \hfill
    \begin{subfigure}[t]{0.32\textwidth}
        \centering
        \includegraphics[width=\textwidth]{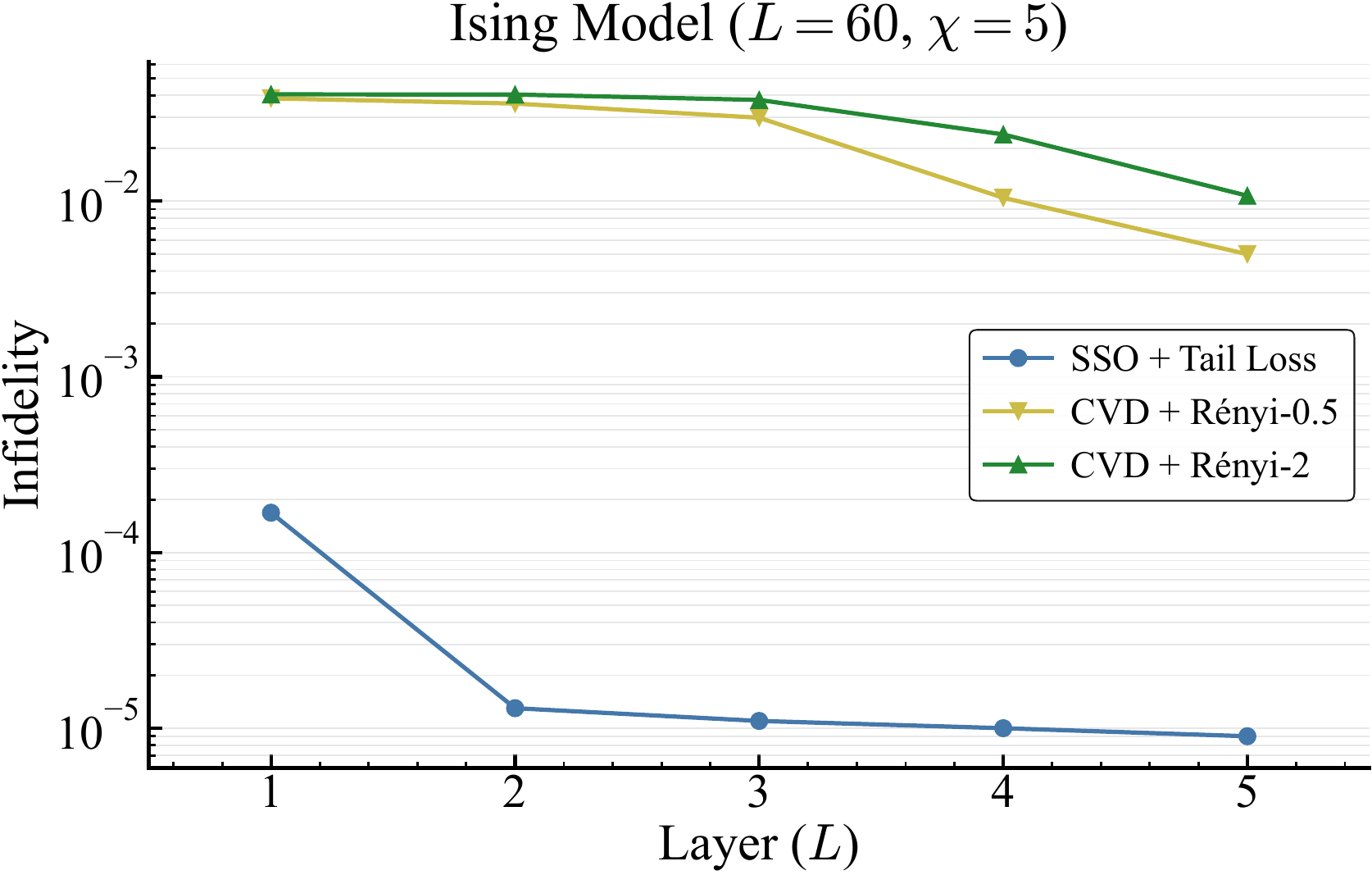}
        \caption{}
        \label{fig:sub3}
    \end{subfigure}
    \caption{
        Comparison of the proposed SSO algorithm with the $\chi=2$ tail loss against
        Classical Variational Disentanglement (CVD)~\cite{mansuroglu2026}
        across three tasks:
        (a)~a random MPS on $n=10$ qubits with maximum bond dimension $\chi=8$;
        (b)~a random MPS on $n=16$ qubits with $\chi=16$; and
        (c)~the ground-state of the $n=60$ tilted spin-$1/2$ Ising chain
        $H = -\sum_i Z_i Z_{i+1} + 0.5 \sum_i X_i + 0.05 \sum_i Z_i$ truncated to a $\chi=5$ MPS target.
        We benchmark against CVD with Rényi-$\alpha$ entropy targets at $\alpha=0.5$
        and $\alpha=2$. Panels~(a) and~(b) additionally report the SSO algorithm with
        a Rényi-2 target, and panel~(a) further includes CVD equipped with the
        $\chi=2$ tail loss introduced in this work. Across all three benchmarks, the SSO algorithm with the
        $\chi=2$ tail loss attains the highest fidelity at every circuit depth.
}
    \label{fig:comparison_with_cvd}
\end{figure*}

We first benchmark against the Classical Variational Disentanglement (CVD) method \cite{mansuroglu2026}. See Fig.~\ref{fig:comparison_with_cvd}. We report fidelity as
\begin{equation}
    F_S=|\braket{\psi_{\text{targ}}|U_S|0^{\otimes n}}|^2=|\braket{\psi_{\text{targ}}|\psi_{\text{prep}}}|^2
\end{equation}
where $U_S$ is the state preparation circuit preparing $\ket{\psi_{\text{prep}}}$. Infidelity is likewise defined as $\epsilon_S=1-F_S$, and is used alongside a log-scale in Fig.~\ref{fig:comparison_with_cvd}(c) for visual clarity. We note that our use of fidelity differs from the metric adopted in \cite{mansuroglu2026}, see Appendix~\ref{sec:fidelity_per_site_cvd_comparison} for details. Layer counts are parameter-matched between methods, with each layer containing $n-1$ 2-qubit gates for an $n$-qubit target state.

In all cases, and most notably for the 60-qubit Ising model target, the SSO algorithm with the tail loss defined in Eq.~\ref{eq:new_cost_fn} achieves higher fidelity encodings compared to the CVD algorithm using Rényi-$\alpha$ entropy with $\alpha\in\{0.5,2\}$. We also experiment with an alternative CVD-inspired Rényi-2 entropy loss function in place of the tail loss in Fig.~\ref{fig:comparison_with_cvd}(a--b), finding that this modestly weakens performance. Interestingly, in Fig.~\ref{fig:comparison_with_cvd}(a) we also find that adopting the SSO-style tail loss in the CVD algorithm significantly deteriorates performance. This demonstrates that the improved performance of the SSO algorithm is not simply explained by the loss function. Rather, the core advantage is derived from the tail loss objective coupled with the final layer being an analytical mapping ensuring that the final fidelity is exactly the $\chi=2$ approximation to the final disentangled MPS (cf. Eq.~\ref{eq:fidelity_equality}). 

In contrast, the CVD algorithm aims to disentangle the target state to a product state entirely with variational operators. This also implies that the worst-case performance of the SSO algorithm is a circuit preparing the $\chi=2$ approximation to the target state, whereas the CVD algorithm lacks this guarantee. For example, the $\chi=2$ approximation to the Ising model target is $F_{\chi=2}=0.99983$, obtained in a single layer by the SSO algorithm. The CVD algorithm exhibits somewhat pathological behaviour here, not reaching this $\chi=2$ approximation fidelity even after 5 layers.

\subsection{Comparison with Matrix Product Disentangler (MPD) and Optimised Variants}

\begin{figure*}[hbtp!]
    \centering

    \begin{minipage}{0.48\textwidth}
        \centering
        \textbf{(i) 1D Quantum Ising}\\[0.3em]
        \includegraphics[width=\linewidth]{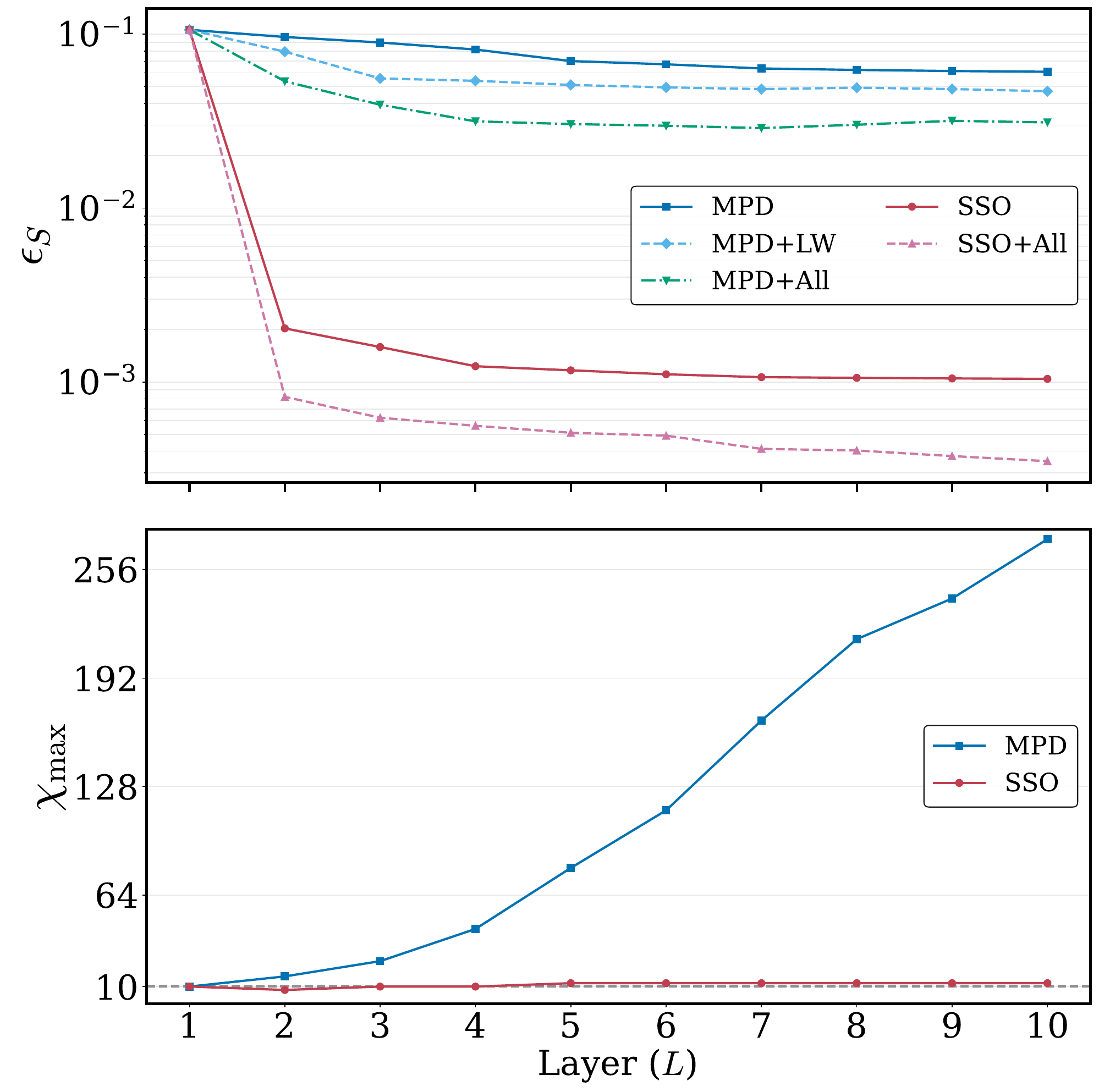}
    \end{minipage}
    \hfill
    \begin{minipage}{0.48\textwidth}
        \centering
        \textbf{(ii) 1D MBL Hamiltonian}\\[0.3em]
        \includegraphics[width=\linewidth]{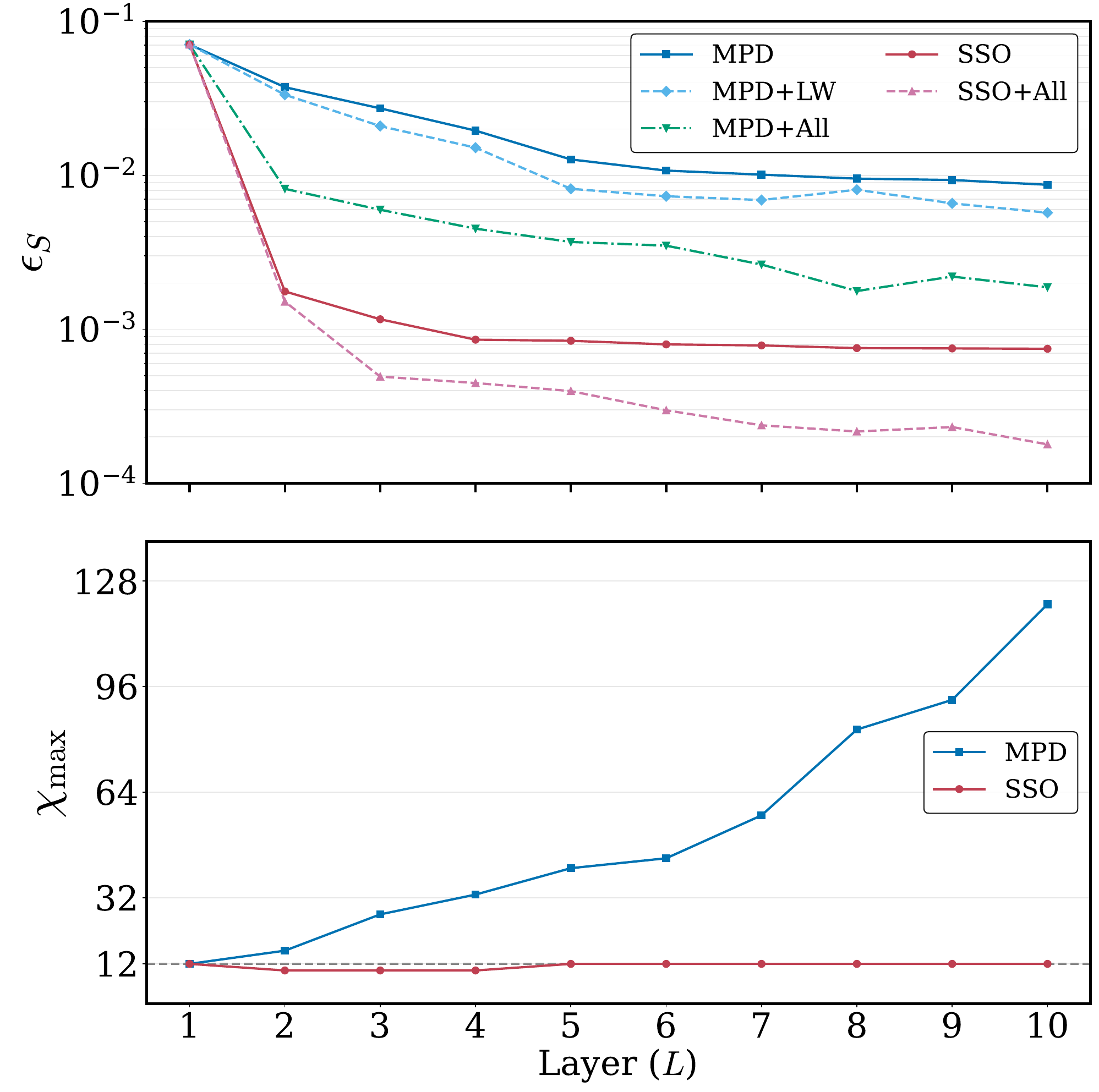}
    \end{minipage}

    \vspace{1em}

    \begin{minipage}{0.48\textwidth}
        \centering
        \textbf{(iii) 1D Hubbard Hamiltonian}\\[0.3em]
        \includegraphics[width=\linewidth]{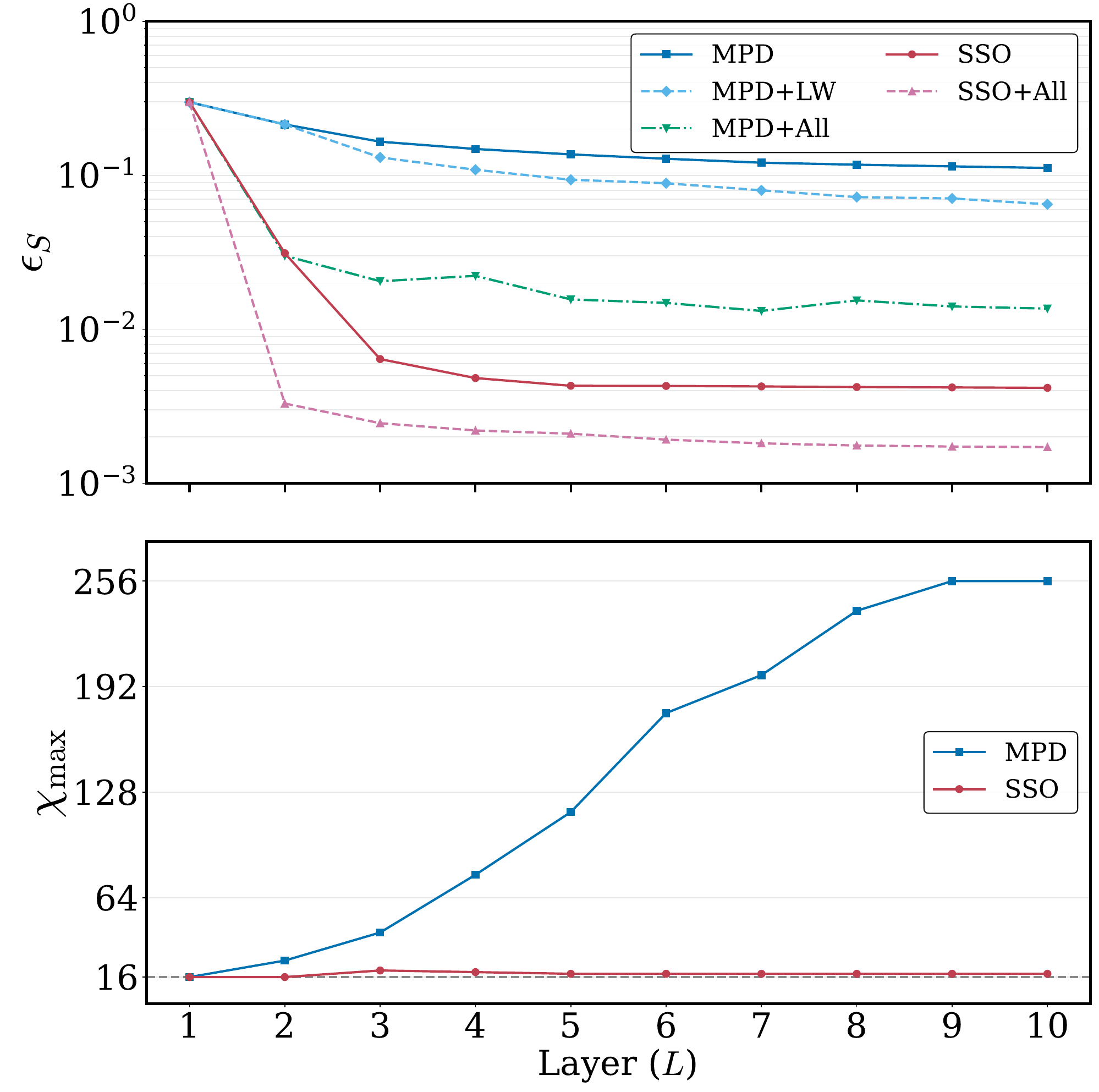}
    \end{minipage}
    \hfill
    \begin{minipage}{0.48\textwidth}
        \centering
        \textbf{(iv) 2D Heisenberg Hamiltonian}\\[0.3em]
        \includegraphics[width=\linewidth]{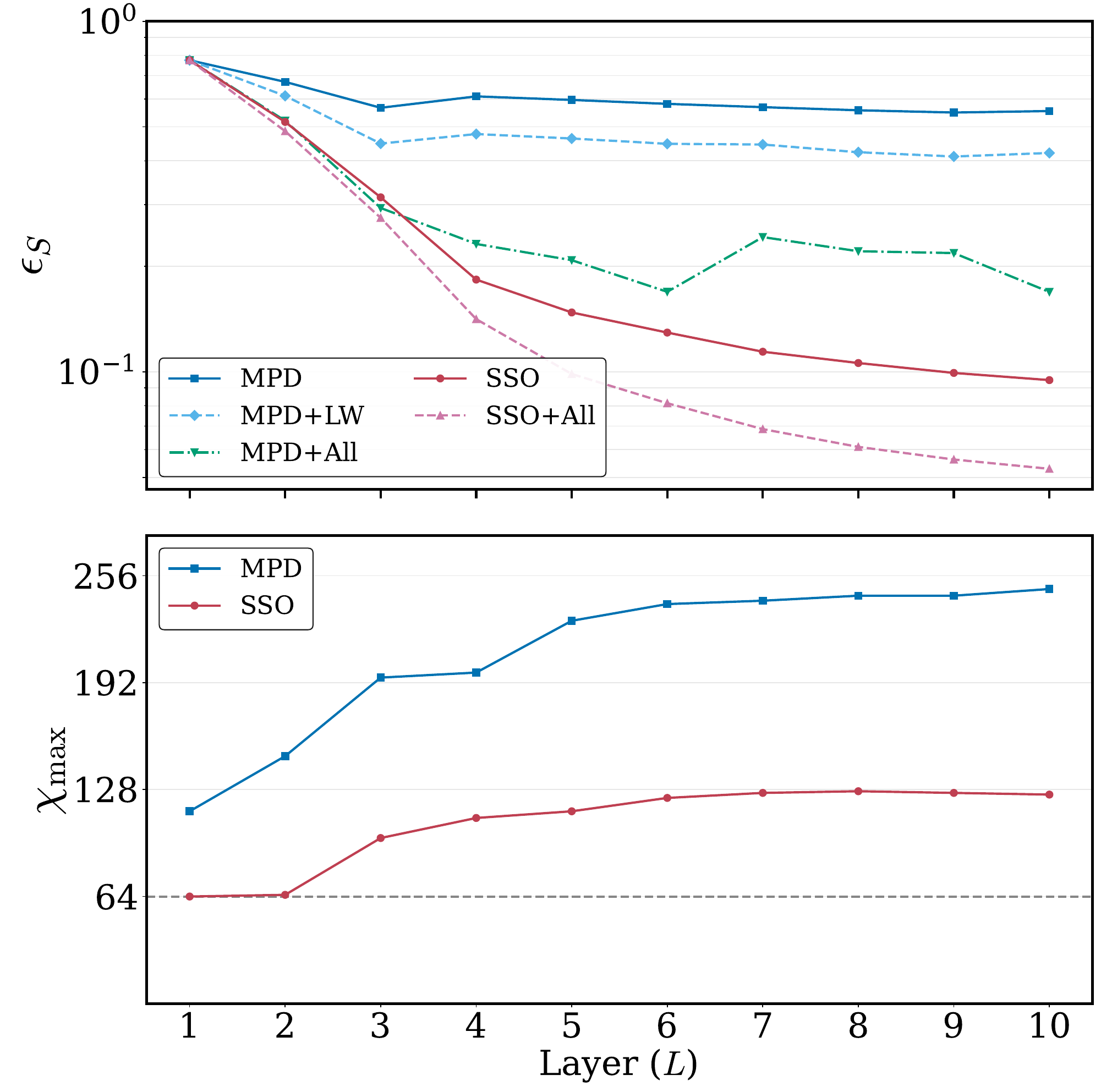}
    \end{minipage}

    \caption{The performance of the SSO algorithm and its post-processed version (SSO+All) with the MPD and its optimsied variants. We test across four MPS approximations of the ground-states of lattice Hamiltonians: (i) The 1D quantum Ising model near-criticality with $n=48$ qubits, (ii) A disordered spin model in the many-body localised regime with $n=16$ qubits, (iii) The spinless Hubbard Chain with $n=16$ qubits, and (iv) The 2D Heisenberg with nearest-neighbour coupling on a $4\times4$ grid of qubits. \textit{Above plot:} the state preparation error $\epsilon_S=1-F_S$ across layers $L$, where $F_S$ is the fidelity between the simulated circuit output and the target MPS. \textit{Below plot:} the scaling of $\chi_{\text{max}}$ during disentangling using $\lambda_{\text{thresh}}=10^{-7}$. Best of 2 runs reported.}
    \label{fig:core_results}
\end{figure*}

We compute MPS approximations to the ground-states of four lattice Hamiltonians via DMRG \cite{White1992DMRG,White1993DMRG}. Here we report the infidelity of the prepared state as
\begin{equation}
    \epsilon_S=1-F_S=1-|\braket{\psi_{\text{targ}}|U_S|0^{\otimes n}}|^2
\end{equation}
where $U_S$ is the state preparation circuit generated by the SSO algorithm. We also report the scaling of the maximum bond dimension $\chi_{\text{max}}$ using a Schmidt coefficient threshold of $\lambda_{\text{thresh}}=10^{-7}$ to truncate the bond dimension of each intermediate MPS after each disentangling layer. For this shallow-depth case study, we do not limit the intermediate bond dimension.

We first benchmark on the 48-site quantum Ising chain in a transverse field,
\begin{equation}
    H = - \sum_{i=1}^{n-1} Z_i Z_{i+1} - h_x \sum_{i=1}^n X_i
\end{equation}
with $n=48$, $J=1.0$, $h_x = 0.5$, and open boundary conditions (i.e. the system is inside the gapped ferromagnetic phase) \cite{Schollwock2011,Pfeuty1970TFIM}. The target bond dimension is $\chi_{\text{Ising}}=10$. We find a significant gap in performance (see Fig.\ref{fig:core_results}(i)): the SSO algorithm achieves a significantly reduced infidelity and no significant increase in $\chi_{\text{max}}$. In this case, the MPD algorithm exhibits pathological behaviour, with $\chi_{\text{max}}$ growing nearly exponentially while the disentangling procedure is largely ineffective. The SSO algorithm achieves a prepared fidelity of $F=0.9980$ within just 2 layers (i.e., only a single \emph{optimised} circuit layer). There is still evidence of local minima, however, with the infidelity curve exhibiting plateauing gains.

Next, we benchmark on target MPS approximations to the ground-states of three other lattice Hamiltonians. As a disordered spin model in the many-body localised regime, we consider
an $n=16$-site spin-$\tfrac{1}{2}$ chain with random longitudinal fields of strength $d h = 1.0$
drawn from a Gaussian distribution (target $\chi_{\text{MBL}}=10$). As a one-dimensional fermionic model, we use the spinless
Hubbard chain with $n=16$ sites, hopping amplitude $t=0.5$, nearest–neighbour interaction
strength $V=1.0$, and chemical potential $\mu=1.0$, with open boundary conditions (target $\chi_{\text{Hubbard}}=16$). Finally, we study the spin-$\tfrac{1}{2}$ 2D Heisenberg model on a $4\times 4$ square lattice with nearest–neighbour coupling $J=1.0$ and open boundary conditions (target $\chi_{\text{Heisenberg}}=64$) \cite{ricardo2005}. 

In all benchmarks, the SSO algorithm reduced error in the prepared state by approximately an order of magnitude compared to the MPD algorithm baseline at fixed circuit depth. In computing more effective disentangling layers, the SSO algorithm also exhibits diminished growth in $\chi_{\text{max}}$, providing evidence of it being a more scalable approach. The SSO algorithm also outperforms the optimised MPD+LW and MPD+All algorithms. Despite these alternative methods also using tensor network optimisation, this is clear evidence that such methods suffer from subpar local minima, and that the SSO algorithm offers an improved optimisation strategy for maximising fidelity at fixed circuit depth.

The SSO+All algorithm performs the best across all test cases. This is unsurprising: since the parameters are initialised by the output of the  SSO algorithm, the \textit{worst-case} fidelity achieved by SSO+All is exactly equal to that obtained by the baseline SSO algorithm. In reality, we find that the joint optimisation of layers often significantly improves outcomes. At deeper layers, however, the joint optimisation plateaus (and sometimes worsens, e.g. see MPD+All in Fig.~\ref{fig:core_results}(iv)). Despite the circuit being deeper, the optimisation process is unable to exploit the additional expressivity of the circuit because the parameter initialisation has only marginally improved, while deeper circuits increase the risk of suboptimal local minima \cite{liu2022}. It is clear from the comparative performance of the MPD+All and SSO+All algorithms that the quality of the parameter initialisation plays a critical role in mitigating the effects of local minima during optimisation.

\section{Discussion and Conclusions}

In this work, we introduced Schmidt Spectrum Optimisation (SSO) as a novel approach to quantum state preparation based on initially disentangling a target state towards a product state. The SSO algorithm computes a circuit consisting entirely of local one- and two-qubit gates, which is optimised entirely by a classical computer. After successfully disentangling the target state, an efficient quantum state preparation circuit is formed by reversing the set of disentangling layers. This protocol works for any target state described by a Matrix Product State (MPS).

The SSO algorithm represents an innovation over existing preparation-via-disentangling approaches \cite{Ran2020,Rudolph2022,BenDov2024,mansuroglu2026}, introducing a novel objective based on minimising the error of the $\chi=2$ truncation to each intermediate MPS in the disentangling process. Across all benchmarks, the SSO algorithm was found to produce the most accurate state preparation circuits when compared to alternative approaches at fixed circuit depth. Further, the SSO algorithm exhibited significantly improved empirical scaling when compared to the MPD algorithm, evidencing it as a scalable approach to state preparation. 

Importantly, the SSO algorithm outperformed the MPD+All algorithm that enacts a joint tensor network optimisation of all circuits layers initialised by the MPD algorithm \cite{Rudolph2022,BenDov2024}. This provides clear evidence that such methods suffer from subpar local minima, even when provided with the parameter initialisation provided by the MPD algorithm. We explored applying a similar joint optimisation of all circuit layers when initialised by the SSO algorithm (SSO+All), which was found to further improve results and was found to be the best performing algorithm across benchmarks. However, a global tensor network optimisation of this form is unlikely to be efficiently scalable beyond shallow-depth circuits. The SSO+All algorithm is nonetheless a promising approach for practical state preparation on near-term quantum hardware. 

The SSO algorithm was found to outperform Classical Variational Disentanglement (CVD) \cite{mansuroglu2026}, which we attribute to an improved design based on aiming to maximise the accuracy of the $\chi=2$ approximation to the final disentangled MPS, before analytically preparing this state with a single additional circuit layer. Further, it significantly improved upon related optimisation techniques (MPD+LW) that iteratively increase entanglement to approximate the target state. We suggest that fidelity-maximising forward approaches like these are constrained by the requirement that the output state after each layer should always approximate the entanglement pattern of the target state. In contrast, optimised preparation-via-disentangling approaches like the CVD and SSO algorithms are alleviated of this constraint, as intermediate disentangled states can take any form as long as they further minimise the relevant loss functions. Such approaches therefore have more freedom during optimisation, allowing for improved outcomes. Nonetheless, the SSO algorithm still exhibited plateauing fidelity curves with increasing depth, suggesting that it still suffers from local minima during optimisation.

Future work may consider generalising this approach to other isometric acyclic tensor network structures, such as binary Tree Tensor Network (bTTN), known to have greater expressive power when compared to the MPS ansatz \cite{Sugawara2025}. While the tail loss function adopted by the SSO algorithm exhibited improved performance over existing methods, it does not rule out the possibility that other loss functions and optimised disentangling designs may find further improvements. Further, our work suggests that the specific relationship between the loss function and ansatz plays an important role. We leave exploration of alternative ansatz for future work. 

In conclusion, the SSO algorithm reframes MPS disentangling as a direct optimisation of the Schmidt spectra at every bond. Across ground-state benchmarks on up to 60 qubits, it exhibits improved accuracy and scalability over existing approaches. Since it produces classically computed quantum circuits consisting of local two-qubit gates, it is particularly well-suited to near-term quantum hardware. We therefore propose SSO as a drop-in replacement for MPD-style disentangling in shallow-depth MPS state preparation.

\appendix

\section{Relation Between Fidelity and Per-Site Error}
\label{sec:fidelity_per_site_cvd_comparison}

Here we clarify an important difference between the fidelity metrics reported in our work and the per-site overlap error metrics introduced in the CVD paper \cite{mansuroglu2026}. Let
\[
F = |\langle \psi_{\mathrm{prep}} | \psi_{\mathrm{target}} \rangle|^2
\]
be the fidelity between $n$-qubit prepared and target states. The per-site overlap error is defined as
\[
\epsilon_{\mathrm{site}}
= 1 - (1-\epsilon)^{1/n}
= 1 - F^{1/(2n)}.
\]
where 
\[
\epsilon = 1 - |\langle \psi_{\mathrm{prep}} | \psi_{\mathrm{target}} \rangle|,
\]
is the global overlap error. Equivalently, given a per-site error \(\epsilon_{\mathrm{site}}\),
the corresponding fidelity is
\[
F = (1-\epsilon_{\mathrm{site}})^{2n}.
\]
We note that the exponential dependence on $n$ can imply stark differences between the scale of metrics, especially for large numbers of qubits. For example, a per-site error of $\epsilon_{\mathrm{site}} = 10^{-4}$ for a 50-qubit target state corresponds to a fidelity of $F\approx0.9900$ (an infidelity of $\epsilon\approx10^{-2}$).

\bibliographystyle{quantum}
\bibliography{references}

@article{mansuroglu2026,
  title = {Preparation circuits for matrix product states by classical variational disentanglement},
  author = {Mansuroglu, Refik and Schuch, Norbert},
  journal = {Phys. Rev. A},
  volume = {113},
  issue = {4},
  pages = {042430},
  numpages = {15},
  year = {2026},
  month = {Apr},
  publisher = {American Physical Society},
  doi = {10.1103/7x2g-twkl},
  url = {https://link.aps.org/doi/10.1103/7x2g-twkl}
}

@article{iaconis2024,
  author    = {Iaconis, Jason and Johri, Sonika and Zhu, Elton Yechao},
  title     = {Quantum state preparation of normal distributions using matrix product states},
  journal   = {npj Quantum Information},
  volume    = {10},
  number    = {1},
  pages     = {15},
  year      = {2024},
  date      = {2024-01-25},
  doi       = {10.1038/s41534-024-00805-0},
  url       = {https://doi.org/10.1038/s41534-024-00805-0},
  ISSN      = {2056-6387},
  publisher = {IOP Publishing Ltd}
}

@article{wei_malz2022,
  title = {Sequential Generation of Projected Entangled-Pair States},
  author = {Wei, Zhi-Yuan and Malz, Daniel and Cirac, J. Ignacio},
  journal = {Phys. Rev. Lett.},
  volume = {128},
  issue = {1},
  pages = {010607},
  numpages = {6},
  year = {2022},
  month = {Jan},
  publisher = {American Physical Society},
  doi = {10.1103/PhysRevLett.128.010607},
  url = {https://link.aps.org/doi/10.1103/PhysRevLett.128.010607}
}

@article{Mottonen2004,
  title = {Quantum Circuits for General Multiqubit Gates},
  author = {M\"ott\"onen, Mikko and Vartiainen, Juha J. and Bergholm, Ville and Salomaa, Martti M.},
  journal = {Phys. Rev. Lett.},
  volume = {93},
  issue = {13},
  pages = {130502},
  numpages = {4},
  year = {2004},
  month = {Sep},
  publisher = {American Physical Society},
  doi = {10.1103/PhysRevLett.93.130502},
  url = {https://link.aps.org/doi/10.1103/PhysRevLett.93.130502}
}

@article{Green2025,
  title = {Quantum encoding of functions and images with matrix product states},
  author = {Green, Josh and Wang, Jingbo},
  journal = {Phys. Rev. A},
  volume = {113},
  issue = {5},
  pages = {052616},
  numpages = {23},
  year = {2026},
  month = {May},
  publisher = {American Physical Society},
  doi = {10.1103/1mn6-j57y},
  url = {https://link.aps.org/doi/10.1103/1mn6-j57y}
}

@Article{Krol2022,
AUTHOR = {Krol, Anna M. and Sarkar, Aritra and Ashraf, Imran and Al-Ars, Zaid and Bertels, Koen},
TITLE = {Efficient Decomposition of Unitary Matrices in Quantum Circuit Compilers},
JOURNAL = {Applied Sciences},
VOLUME = {12},
YEAR = {2022},
NUMBER = {2},
ARTICLE-NUMBER = {759},
URL = {https://www.mdpi.com/2076-3417/12/2/759},
ISSN = {2076-3417},
DOI = {10.3390/app12020759}
}

@article{vidal2008,
  title = {Class of Quantum Many-Body States That Can Be Efficiently Simulated},
  author = {Vidal, G.},
  journal = {Phys. Rev. Lett.},
  volume = {101},
  issue = {11},
  pages = {110501},
  numpages = {4},
  year = {2008},
  month = {Sep},
  publisher = {American Physical Society},
  doi = {10.1103/PhysRevLett.101.110501},
  url = {https://link.aps.org/doi/10.1103/PhysRevLett.101.110501}
}

@article{melnikov2023,
  author    = {Melnikov, Ar and Termanova, Alena and Dolgov, S and Neukart, Florian and Perelshtein, M},
  title     = {Quantum state preparation using tensor networks},
  journal   = {Quantum Science and Technology},
  volume    = {8},
  number    = {3},
  pages     = {035027},
  year      = {2023},
  month     = {06},
  date      = {2023-06-19},
  doi       = {10.1088/2058-9565/acd9e7},
  url       = {https://iopscience.iop.org/article/10.1088/2058-9565/acd9e7},
  publisher = {IOP Publishing Ltd}
}

@misc{Sugawara2025,
      title={Embedding of Tree Tensor Networks into Shallow Quantum Circuits}, 
      author={Shota Sugawara and Kazuki Inomata and Tsuyoshi Okubo and Synge Todo},
      year={2025},
      eprint={2501.18856},
      archivePrefix={arXiv},
      primaryClass={quant-ph},
      url={https://arxiv.org/abs/2501.18856}, 
}

@article{Schuhmacher_2025,
   title={Hybrid Tree Tensor Networks for Quantum Simulation},
   volume={6},
   ISSN={2691-3399},
   url={http://dx.doi.org/10.1103/PRXQuantum.6.010320},
   DOI={10.1103/prxquantum.6.010320},
   number={1},
   journal={PRX Quantum},
   publisher={American Physical Society (APS)},
   author={Schuhmacher, Julian and Ballarin, Marco and Baiardi, Alberto and Magnifico, Giuseppe and Tacchino, Francesco and Montangero, Simone and Tavernelli, Ivano},
   year={2025},
   month=jan }

@article{yu2024,
  title = {Dual-Isometric Projected Entangled Pair States},
  author = {Yu, Xie-Hang and Cirac, J. Ignacio and Kos, Pavel and Styliaris, Georgios},
  journal = {Phys. Rev. Lett.},
  volume = {133},
  issue = {19},
  pages = {190401},
  numpages = {7},
  year = {2024},
  month = {Nov},
  publisher = {American Physical Society},
  doi = {10.1103/PhysRevLett.133.190401},
  url = {https://link.aps.org/doi/10.1103/PhysRevLett.133.190401}
}

@article{cincio2008,
  title = {Multiscale Entanglement Renormalization Ansatz in Two Dimensions: Quantum Ising Model},
  author = {Cincio, Lukasz and Dziarmaga, Jacek and Rams, Marek M.},
  journal = {Phys. Rev. Lett.},
  volume = {100},
  issue = {24},
  pages = {240603},
  numpages = {4},
  year = {2008},
  month = {Jun},
  publisher = {American Physical Society},
  doi = {10.1103/PhysRevLett.100.240603},
  url = {https://link.aps.org/doi/10.1103/PhysRevLett.100.240603}
}

@article{yu2017,
  title = {Finding Matrix Product State Representations of Highly Excited Eigenstates of Many-Body Localized Hamiltonians},
  author = {Yu, Xiongjie and Pekker, David and Clark, Bryan K.},
  journal = {Phys. Rev. Lett.},
  volume = {118},
  issue = {1},
  pages = {017201},
  numpages = {5},
  year = {2017},
  month = {Jan},
  publisher = {American Physical Society},
  doi = {10.1103/PhysRevLett.118.017201},
  url = {https://link.aps.org/doi/10.1103/PhysRevLett.118.017201}
}

@article{Jaderberg2025,
  title = {Variational preparation of normal matrix product states on quantum computers},
  author = {Jaderberg, Ben and Pennington, George and Marshall, Kate V. and Anderson, Lewis W. and Agarwal, Abhishek and Lindoy, Lachlan P. and Rungger, Ivan and Mensa, Stefano and Crain, Jason},
  journal = {Phys. Rev. Res.},
  pages = {--},
  year = {2025},
  month = {Dec},
  publisher = {American Physical Society},
  doi = {10.1103/m5kb-4f43},
  url = {https://link.aps.org/doi/10.1103/m5kb-4f43}
}

@article{Cirac2021,
  title = {Matrix product states and projected entangled pair states: Concepts, symmetries, theorems},
  author = {Cirac, J. Ignacio and P\'erez-Garc\'{\i}a, David and Schuch, Norbert and Verstraete, Frank},
  journal = {Rev. Mod. Phys.},
  volume = {93},
  issue = {4},
  pages = {045003},
  numpages = {65},
  year = {2021},
  month = {Dec},
  publisher = {American Physical Society},
  doi = {10.1103/RevModPhys.93.045003},
  url = {https://link.aps.org/doi/10.1103/RevModPhys.93.045003}
}

@article{Schwarz2012,
  title = {Preparing Projected Entangled Pair States on a Quantum Computer},
  author = {Schwarz, Martin and Temme, Kristan and Verstraete, Frank},
  journal = {Phys. Rev. Lett.},
  volume = {108},
  issue = {11},
  pages = {110502},
  numpages = {5},
  year = {2012},
  month = {Mar},
  publisher = {American Physical Society},
  doi = {10.1103/PhysRevLett.108.110502},
  url = {https://link.aps.org/doi/10.1103/PhysRevLett.108.110502}
}

@article{Manabe2025,
  doi = {10.22331/q-2025-05-28-1755},
  url = {https://doi.org/10.22331/q-2025-05-28-1755},
  title = {The {S}tate {P}reparation of {M}ultivariate {N}ormal {D}istributions using {T}ree {T}ensor {N}etwork},
  author = {Manabe, Hidetaka and Sano, Yuichi},
  journal = {{Quantum}},
  issn = {2521-327X},
  publisher = {{Verein zur F{\"{o}}rderung des Open Access Publizierens in den Quantenwissenschaften}},
  volume = {9},
  pages = {1755},
  month = may,
  year = {2025}
}

@article{Ran2020,
  author  = {Ran, Shi-Ju},
  title   = {Encoding of matrix product states into quantum circuits of one- and two-qubit gates},
  journal = {Phys. Rev. A},
  volume  = {101},
  number  = {3},
  pages   = {032310},
  year    = {2020},
  doi     = {10.1103/PhysRevA.101.032310}
}

@article{Rudolph2022,
  title        = {Decomposition of matrix product states into shallow quantum circuits},
  author       = {Rudolph, Manuel S and Chen, Jing and Miller, Jacob and Acharya, Atithi and Perdomo-Ortiz, Alejandro},
  journal      = {Quantum Science and Technology},
  volume       = {9},
  number       = {1},
  pages        = {015012},
  year         = {2024},
  doi          = {10.1088/2058-9565/ad04e6},
  url          = {https://doi.org/10.1088/2058-9565/ad04e6}
}

@article{Eisert2010AreaLaws,
  title        = {Colloquium: Area laws for the entanglement entropy},
  author       = {Eisert, Jens and Cramer, Marcus and Plenio, Martin B.},
  journal      = {Reviews of Modern Physics},
  volume       = {82},
  number       = {1},
  pages        = {277--306},
  year         = {2010},
  month        = feb,
  doi          = {10.1103/RevModPhys.82.277},
  url          = {https://doi.org/10.1103/RevModPhys.82.277}
}

@article{WhiteScalapino1998,
  author  = {White, Steven R. and Scalapino, Douglas J.},
  title   = {Density matrix renormalization group study of the striped phase in the 2D Hubbard model},
  journal = {Phys. Rev. Lett.},
  volume  = {80},
  pages   = {1272--1275},
  year    = {1998},
  doi     = {10.1103/PhysRevLett.80.1272}
}

@article{BenDov2024,
  title = {Approximate encoding of quantum states using shallow circuits},
  author = {Ben‑Dov, Matan and Shnaiderov, David and Makmal, Adi and Dalla Torre, Emanuele G.},
  journal = {npj Quantum Information},
  volume = {10},
  number = {1},
  pages = {65},
  year = {2024},
  doi = {10.1038/s41534-024-00858-1}
}

@article{Malz2024,
  title = {Preparation of Matrix Product States with Log‑Depth Quantum Circuits},
  author = {Malz, Daniel and Styliaris, Georgios and Wei, Zhi‑Yuan and Cirac, J. Ignacio},
  journal = {Physical Review Letters},
  volume = {132},
  number = {4},
  pages = {040404},
  year = {2024},
  doi = {10.1103/PhysRevLett.132.040404}
}

@article{schon2005,
  title = {Sequential Generation of Entangled Multiqubit States},
  author = {Sch\"on, C. and Solano, E. and Verstraete, F. and Cirac, J. I. and Wolf, M. M.},
  journal = {Phys. Rev. Lett.},
  volume = {95},
  issue = {11},
  pages = {110503},
  numpages = {4},
  year = {2005},
  month = {Sep},
  publisher = {American Physical Society},
  doi = {10.1103/PhysRevLett.95.110503},
  url = {https://link.aps.org/doi/10.1103/PhysRevLett.95.110503}
}

@article{Smith2024,
  title = {Constant‑Depth Preparation of Matrix Product States with Adaptive Quantum Circuits},
  author = {Smith, Kevin C. and Khan, Abid and Clark, Bryan K. and Girvin, S.M. and Wei, Tzu‑Chieh},
  journal = {PRX Quantum},
  volume = {5},
  number = {3},
  pages = {030344},
  year = {2024},
  doi = {10.1103/PRXQuantum.5.030344}
}

@article{White1992DMRG,
  title   = {Density Matrix Formulation for Quantum Renormalization Groups},
  author  = {White, Steven R.},
  journal = {Physical Review Letters},
  volume  = {69},
  number  = {19},
  pages   = {2863--2866},
  year    = {1992},
  doi     = {10.1103/PhysRevLett.69.2863}
}

@article{Jobst2024efficientmps,
  doi = {10.22331/q-2024-12-03-1544},
  url = {https://doi.org/10.22331/q-2024-12-03-1544},
  title = {Efficient {MPS} representations and quantum circuits from the {F}ourier modes of classical image data},
  author = {Jobst, Bernhard and Shen, Kevin and Riofr{\'{i}}o, Carlos A. and Shishenina, Elvira and Pollmann, Frank},
  journal = {{Quantum}},
  issn = {2521-327X},
  publisher = {{Verein zur F{\"{o}}rderung des Open Access Publizierens in den Quantenwissenschaften}},
  volume = {8},
  pages = {1544},
  month = dec,
  year = {2024}
}

@article{dong2022,
  title = {Ground-State Preparation and Energy Estimation on Early Fault-Tolerant Quantum Computers via Quantum Eigenvalue Transformation of Unitary Matrices},
  author = {Dong, Yulong and Lin, Lin and Tong, Yu},
  journal = {PRX Quantum},
  volume = {3},
  issue = {4},
  pages = {040305},
  numpages = {25},
  year = {2022},
  month = {Oct},
  publisher = {American Physical Society},
  doi = {10.1103/PRXQuantum.3.040305},
  url = {https://link.aps.org/doi/10.1103/PRXQuantum.3.040305}
}

@article{Bausch2022fastblackboxquantum,
  doi = {10.22331/q-2022-08-04-773},
  url = {https://doi.org/10.22331/q-2022-08-04-773},
  title = {Fast {B}lack-{B}ox {Q}uantum {S}tate {P}reparation},
  author = {Bausch, Johannes},
  journal = {{Quantum}},
  issn = {2521-327X},
  publisher = {{Verein zur F{\"{o}}rderung des Open Access Publizierens in den Quantenwissenschaften}},
  volume = {6},
  pages = {773},
  month = aug,
  year = {2022}
}

@article{huggins2025,
  title = {Efficient State Preparation for the Quantum Simulation of Molecules in First Quantization},
  author = {Huggins, William J. and Leimkuhler, Oskar and Stetina, Torin F. and Whaley, K. Birgitta},
  journal = {PRX Quantum},
  volume = {6},
  issue = {2},
  pages = {020319},
  numpages = {54},
  year = {2025},
  month = {Apr},
  publisher = {American Physical Society},
  doi = {10.1103/PRXQuantum.6.020319},
  url = {https://link.aps.org/doi/10.1103/PRXQuantum.6.020319}
}

@article{Rosenkranz2025quantumstate,
  doi = {10.22331/q-2025-04-09-1703},
  url = {https://doi.org/10.22331/q-2025-04-09-1703},
  title = {Quantum state preparation for multivariate functions},
  author = {Rosenkranz, Matthias and Brunner, Eric and Marin-Sanchez, Gabriel and Fitzpatrick, Nathan and Dilkes, Silas and Tang, Yao and Kikuchi, Yuta and Benedetti, Marcello},
  journal = {{Quantum}},
  issn = {2521-327X},
  publisher = {{Verein zur F{\"{o}}rderung des Open Access Publizierens in den Quantenwissenschaften}},
  volume = {9},
  pages = {1703},
  month = apr,
  year = {2025}
}

@article{zhang2022,
  title = {Quantum State Preparation with Optimal Circuit Depth: Implementations and Applications},
  author = {Zhang, Xiao-Ming and Li, Tongyang and Yuan, Xiao},
  journal = {Phys. Rev. Lett.},
  volume = {129},
  issue = {23},
  pages = {230504},
  numpages = {6},
  year = {2022},
  month = {Nov},
  publisher = {American Physical Society},
  doi = {10.1103/PhysRevLett.129.230504},
  url = {https://link.aps.org/doi/10.1103/PhysRevLett.129.230504}
}

@article{OBrien2025quantumstate,
  doi = {10.22331/q-2025-07-03-1786},
  url = {https://doi.org/10.22331/q-2025-07-03-1786},
  title = {Quantum state preparation via piecewise {QSVT}},
  author = {O'Brien, Oliver and S{\"{u}}nderhauf, Christoph},
  journal = {{Quantum}},
  issn = {2521-327X},
  publisher = {{Verein zur F{\"{o}}rderung des Open Access Publizierens in den Quantenwissenschaften}},
  volume = {9},
  pages = {1786},
  month = jul,
  year = {2025}
}

@article{ricardo2005,
  title = {Two-dimensional quantum spin-$\frac{1}{2}$ Heisenberg model with competing interactions},
  author = {Ricardo de Sousa, J. and Branco, N. S.},
  journal = {Phys. Rev. B},
  volume = {72},
  issue = {13},
  pages = {134421},
  numpages = {4},
  year = {2005},
  month = {Oct},
  publisher = {American Physical Society},
  doi = {10.1103/PhysRevB.72.134421},
  url = {https://link.aps.org/doi/10.1103/PhysRevB.72.134421}
}

@article{White1993DMRG,
  title   = {Density-matrix algorithms for quantum renormalization groups},
  author  = {White, Steven R.},
  journal = {Physical Review B},
  volume  = {48},
  number  = {14},
  pages   = {10345--10356},
  year    = {1993},
  doi     = {10.1103/PhysRevB.48.10345}
}

@article{Pfeuty1970TFIM,
  title   = {The one-dimensional Ising model with a transverse field},
  author  = {Pfeuty, Pierre},
  journal = {Annals of Physics},
  volume  = {57},
  number  = {1},
  pages   = {79--90},
  year    = {1970},
  doi     = {10.1016/0003-4916(70)90270-8}
}

@article{Schollwock2011,
  title   = {The density-matrix renormalization group in the age of matrix product states},
  author  = {Schollw{\"o}ck, Ulrich},
  journal = {Annals of Physics},
  volume  = {326},
  pages   = {96--192},
  year    = {2011},
  doi     = {10.1016/j.aop.2010.09.012}
}

@article{Oseledets2011,
  title   = {Tensor-train decomposition},
  author  = {Oseledets, Ivan V.},
  journal = {SIAM Journal on Scientific Computing},
  volume  = {33},
  number  = {5},
  pages   = {2295--2317},
  year    = {2011},
  doi     = {10.1137/090752286}
}

@article{EckartYoung1936,
  title   = {The approximation of one matrix by another of lower rank},
  author  = {Eckart, Carl and Young, Gale},
  journal = {Psychometrika},
  volume  = {1},
  number  = {3},
  pages   = {211--218},
  year    = {1936},
  doi     = {10.1007/BF02288367}
}

@article{ehlers2017,
  title = {Hybrid-space density matrix renormalization group study of the doped two-dimensional Hubbard model},
  author = {Ehlers, G. and White, S. R. and Noack, R. M.},
  journal = {Phys. Rev. B},
  volume = {95},
  issue = {12},
  pages = {125125},
  numpages = {14},
  year = {2017},
  month = {Mar},
  publisher = {American Physical Society},
  doi = {10.1103/PhysRevB.95.125125},
  url = {https://link.aps.org/doi/10.1103/PhysRevB.95.125125}
}

@article{stoudenmire2012,
   author = "Stoudenmire, E.M. and White, Steven R.",
   title = "Studying Two-Dimensional Systems with the Density Matrix Renormalization Group", 
   journal= "Annual Review of Condensed Matter Physics",
   year = "2012",
   volume = "3",
   number = "Volume 3, 2012",
   pages = "111-128",
   doi = "https://doi.org/10.1146/annurev-conmatphys-020911-125018",
   url = "https://www.annualreviews.org/content/journals/10.1146/annurev-conmatphys-020911-125018",
   publisher = "Annual Reviews",
   issn = "1947-5462",
   type = "Journal Article",
  }

@article{verstraete2006,
  title = {Matrix product states represent ground states faithfully},
  author = {Verstraete, F. and Cirac, J. I.},
  journal = {Phys. Rev. B},
  volume = {73},
  issue = {9},
  pages = {094423},
  numpages = {8},
  year = {2006},
  month = {Mar},
  publisher = {American Physical Society},
  doi = {10.1103/PhysRevB.73.094423},
  url = {https://link.aps.org/doi/10.1103/PhysRevB.73.094423}
}

@article{Grant2019,
   title={An initialization strategy for addressing barren plateaus in parametrized quantum circuits},
   volume={3},
   ISSN={2521-327X},
   url={http://dx.doi.org/10.22331/q-2019-12-09-214},
   DOI={10.22331/q-2019-12-09-214},
   journal={Quantum},
   publisher={Verein zur Forderung des Open Access Publizierens in den Quantenwissenschaften},
   author={Grant, Edward and Wossnig, Leonard and Ostaszewski, Mateusz and Benedetti, Marcello},
   year={2019},
   month=dec, pages={214} }

@inproceedings{pytorch,
  author = {Paszke, Adam and Gross, Sam and Chintala, Soumith and Chanan, Gregory and Yang, Edward and DeVito, Zachary and Lin, Zeming and Desmaison, Alban and Antiga, Luca and Lerer, Adam},
  booktitle = {NIPS 2017 Workshop on Autodiff},
  location = {Long Beach, California, USA},
  title = {Automatic Differentiation in PyTorch},
  url = {https://openreview.net/forum?id=BJJsrmfCZ},
  year = 2017
}

@article{Styliaris2025,
   title={Matrix-product unitaries: Beyond quantum cellular automata},
   volume={9},
   ISSN={2521-327X},
   url={http://dx.doi.org/10.22331/q-2025-02-25-1645},
   DOI={10.22331/q-2025-02-25-1645},
   journal={Quantum},
   publisher={Verein zur Forderung des Open Access Publizierens in den Quantenwissenschaften},
   author={Styliaris, Georgios and Trivedi, Rahul and Perez-Garcia, David and Cirac, J. Ignacio},
   year={2025},
   month=feb, pages={1645} }

@article{Mirsky1960,
  title   = {Symmetric gauge functions and unitarily invariant norms},
  author  = {Mirsky, L.},
  journal = {Quarterly Journal of Mathematics},
  volume  = {11},
  number  = {1},
  pages   = {50--59},
  year    = {1960},
  doi     = {10.1093/qmath/11.1.50}
}

@article{liu2022,
  title = {Presence and Absence of Barren Plateaus in Tensor-Network Based Machine Learning},
  author = {Liu, Zidu and Yu, Li-Wei and Duan, L.-M. and Deng, Dong-Ling},
  journal = {Phys. Rev. Lett.},
  volume = {129},
  issue = {27},
  pages = {270501},
  numpages = {8},
  year = {2022},
  month = {Dec},
  publisher = {American Physical Society},
  doi = {10.1103/PhysRevLett.129.270501},
  url = {https://link.aps.org/doi/10.1103/PhysRevLett.129.270501}
}

@article{orus2014,
title = {A practical introduction to tensor networks: Matrix product states and projected entangled pair states},
journal = {Annals of Physics},
volume = {349},
pages = {117-158},
year = {2014},
issn = {0003-4916},
doi = {https://doi.org/10.1016/j.aop.2014.06.013},
url = {https://www.sciencedirect.com/science/article/pii/S0003491614001596},
author = {Román Orús}
}


\end{document}